\documentclass[a4paper]{jpconf}
\usepackage{graphicx}
\usepackage{xspace}     
\usepackage {amsmath}
\usepackage {bm}  
\usepackage[usenames,dvipsnames]{color}
\usepackage{hyperref}
\newcommand{\pT}{\mbox{$p_T$}\xspace}

\newcommand{\Npart}{\mbox{$N_{\rm part}$}\xspace}

\newcommand{\Nch}{\mbox{$N_{\rm ch}$}\xspace}
\newcommand{\Et}{\mbox{${\rm E}_T$}\xspace}

\newcommand{\sqs}{\mbox{$\sqrt{s}$}\xspace}
\newcommand{\sqsn}{\mbox{$\sqrt{s_{_{NN}}}$}\xspace}

\def\lsim{\raise0.3ex\hbox{$<$\kern-0.75em\raise-1.1ex\hbox{$\sim$}}}
\def\gsim{\raise0.3ex\hbox{$>$\kern-0.75em\raise-1.1ex\hbox{$\sim$}}}

\def\mean#1{\langle #1 \rangle}

\def\QGP{{\color{Red} Q}{\color{Blue} G}{\color{Green} P}}
\def\QCD{{\color{Red} Q}{\color{Green} C}{\color{Blue} D}}
\begin{document}
\title{Is $\mathbf{\hat{q}}$ a physical quantity or just a parameter?\\
and other unanswered questions in high-$\mathbf{\pT}$ physics}

	\author{M.~J.~Tannenbaum\footnote{Supported by the U.S. Department of Energy, Contracts DE-AC02-98CH10886 and DE-SC0012704.}}

\address{Physics Department, Brookhaven National Laboratory, Upton, NY 11973-5000 USA}

\ead{mjt@bnl.gov}

\begin{abstract}
The many different theoretical studies of energy loss of a quark or gluon traversing a medium have one thing in common: the transport coefficient of a gluon in the medium, denoted $\hat{q}$, which is defined as the mean 4-momentum transfer-square,  $q^2$, by a gluon to the medium per gluon mean free path, $\lambda_{\rm mfp}$. In the original BDMPSZ formalism, the energy loss of an outgoing parton, $-dE/dx$,  
per unit length ($x$) of a medium with total length $L$, due to coherent gluon bremsstrahlung is proportional to the $q^2$ and takes the form:
\begin{equation}
{-dE \over dx }\simeq \alpha_s \langle{q^2(L)}\rangle=\alpha_s\, \mu^2\, L/\lambda_{\rm mfp} 
=\alpha_s\, \hat{q}\, L\qquad ,
\end {equation}
where $\mu$, is the mean momentum transfer per collision. Thus, the total energy loss in the medium goes like $L^2$. 

Additionally, the accumulated momentum-square, $\langle{k_{T}^2}\rangle$, transverse to a gluon traversing a length $L$ in the medium is well approximated by $\langle{k_{T}^2}\rangle\approx\langle{q^2(L)}\rangle=\hat{q}\, L$. 
A simple estimate shows that the $\langle{k_{T}^2}\rangle\approx\hat{q}\,L$ should be observable at RHIC at \sqsn=200 GeV via the broadening of di-hadron azimuthal correlations resulting in an azimuthal width $\sim\sqrt{2}$ larger in Au$+$Au   than in $p+p$  collisions . 
Measurements relevant to this issue will be discussed as well as recent STAR jet results presented at QM2014~\cite{YJLeeQM2014}.

Other topics to be discussed include the danger of using forward energy to define centrality in $p(d)+$A collisions for high \pT measurements, the danger of not using comparison $p+p$  data at the same \sqs in the same detector for $R_{\rm{AA}}$ or lately for $R_{\rm{pA}}$ measurements. Also, based on a comment at last year's 9th workshop that the parton energy loss is proportional to $d\Nch/d\eta$~\cite{ShuryaktoMJT2013}, new results on the dependence of the shift in the \pT spectra in A+A collisions from the $T_{\rm{AA}}$-scaled $p+p$  spectrum (to be discussed in detail in another presentation~\cite{TakaoNantes}) will be shown.
\end{abstract}
\section{Introduction--BDMPSZ, the first \QCD\ based Jet Quenching Model}
I don't want to discuss models in detail, since they are nothing like QED or \QCD---theories that you can set your watch by (at least QED). I concentrate on one example, the first \QCD\ based model~\cite{BDMPSZ} which stimulated the use of hard-probes at RHIC as a signature of the \QGP.

It is important to note that the original STAR Letter of Intent (LBL-29651) in 1990, following Wang and Gyulassy (LBL-29390), did cite as one objective: ``the use of hard scattering of partons as a probe of high density nuclear matter... Passage through hadronic or nuclear matter is predicted to result in an attenuation of the jet energy and broadening of jets. Relative to this damped case, a \QGP\ is transparent and an enhanced yield is expected.''

Of course this is precisely the opposite of what was actually discovered at RHIC. 
Furthermore, what had been observed in A+A and $p+$A collisions was an enhancement of the hard scattering, a.k.a. the Cronin Effect~\cite{CroninEffect}, rather than an attenuation. Thus, until the appearance of the fully \QCD\ based models, starting with BDMPS~\cite{BDMPS2}, I described the original Pl\"umer-Gyulassy-Wang~\cite{GyulassyPlumerPLB243,WangGyulassyPRL68} Jet Quenching as ``the vanishing of something that doesn`t exist in the first place'', namely the attenuation of hard-scattering in dense but confined nuclear matter (CNM). 

In the early c.~1990 publications~\cite{GyulassyPlumerPLB243,ThomaGyulassyNPB351} the \QGP\ effect was thought to be ``a sudden {\em decrease} of $dE/dx$ near the quark-gluon plasma phase transition'' which could {\em reduce} the CNM Jet Quenching (``unquench the jets''~\cite{ThomaGyulassyNPB351}) and thus be a possible signature of the \QGP. This idea was downplayed between the original STAR letter of intent in September 1990 and the update in July 1991 as reflected in the new goal for Parton Physics: ``For example, it has been suggested that there will be observable changes in the energy loss of propagating partons as the energy density of the medium increases, particularly if the medium passes through a phase transition to the \QGP~\cite{GyulassyNPA538}''. 

The reason for the downplaying of jet quenching as a possible probe of the \QGP\ by STAR in 1991 was the discovery~\cite{GyulassyNPA538} that instead of small $dE/dx$ in the \QGP\ it was ``recently found that at least deep in the \QGP\ phase, the induced radiative energy loss could be quite large''~\cite{GyulassyNPA538}. Subsequent work found that the radiation was suppressed by the LPM effect~\cite{GyulassyWangNPB420} which led to a series of developments, nicely reviewed in Ref.~\cite{BDMPSZ}, that eventually led to the BDMPSZ \QCD\ based model~\cite{BDMPSZ}. 

\subsection{The real \QGP\ jet quenching and the importance of attending conferences}
I first heard about the original CNM Jet Quenching~\cite{GyulassyPlumerPLB243} at an excellent meeting in Strasbourg in October 1990~\cite{QGPSignaturesStrasbourg} to discuss ``Quark-Gluon Plasma Signatures'' in a talk by Michael Pl\"umer that  was greeted with disbelief by the many CERN-ISR veterans who had puzzled over the Cronin effect for many years. This led to my description noted a few paragraphs above. Meanwhile, the RHIC experiments and ALICE at the LHC~\cite{ALICELOI} were designed with a focus on $J/\Psi$ suppression~\cite{MatsuiSatzPLB178} as the gold-plated signature for deconfinement and the  \QGP.  

In 1998 at the \QCD\ workshop in Paris~\cite{4thQCDWks}, I found what I thought was a cleaner signal of the \QGP\ when Rolf Baier asked me whether jets could be measured in Au$+$Au collisions because he had made studies in p\QCD~\cite{BDMPS2} of the energy loss of partons, produced by hard-scattering ``with their color charge fully exposed'',  in traversing a medium ``with a large density of similarly exposed color charges''. The conclusion was that ``Numerical estimates of the loss suggest that it may be significantly greater in hot matter than in cold. {\em This makes the magnitude of the radiative energy loss a remarkable signal for \QGP\  formation}''~\cite{BDMPSZ}. In addition to being a probe of the \QGP, the fully exposed color charges allow the study of parton-scattering with $Q^2 \ll 1-5$ (GeV/c)$^2$ in the medium where new collective \QCD\ effects may possibly be observed.

Because the expected energy in a typical jet cone $R=\sqrt{(\Delta\eta)^2+ (\Delta\phi)^2}$ in central Au$+$Au collisions at \sqsn=200 GeV would be $\pi R^2\times1/2\pi \times d\Et/d\eta=R^2/2 \times d\Et/d\eta~\sim 300$ GeV for $R=1$, where the kinematic limit is 100 GeV, I said (and wrote~\cite{4thQCDWks}\footnote{It was an excellent guess because the measured $d\Et/d\eta=606\pm 32$ GeV in central Au$+$Au at \sqsn=200\ GeV~\cite{ppg019}.}) that jets can not be reconstructed in Au$+$Au central collisions at RHIC---still correct after 16 years. On the other hand, hard-scattering was discovered in $p+p$ collisions at the CERN-ISR in 1972 with single particle and two-particle correlations, while jets had a long learning curve from 1977--1982 with a notorious false claim (e.g. see Refs.~\cite{RTbook,MJTIJMPA2014}), so I said (and wrote~\cite{4thQCDWks}) that we should use single and two-particle measurements at RHIC---which we did and it WORKED! The present solution for jets in A$+$A collisions (LHC 2010 and RHIC c.2014) is to take smaller cones, with 100 GeV in $R=0.58$, 48 GeV in $R=0.4$, 27 GeV in $R=0.3$, 12 GeV in $R=0.2$ at RHIC.  
\subsection{ ${\hat{q}}$ or di-jet broadening and gluon radiation}
There are many different theoretical studies of energy loss of a quark or gluon with their color charges fully exposed passing through a medium with a large density of similarly exposed color charges (i.e. a \QGP ). The approaches are different, but the one thing that they have in common~\cite{JETcollabPRC90} is the transport coefficient of a gluon in the medium, denoted $\hat{q}$, which is defined as the mean 4-momentum transfer-square, $q^2$, by a gluon to the medium per gluon mean free path, $\lambda_{\rm mfp}$. Thus the mean 4-momentum transfer-square for a gluon traversing length $L$ in the medium is, 
$\mean{q^2(L)}=\hat{q}\,L=\mu^2\,L/\lambda_{\rm mfp}$, where $\mu$, the mean momentum transfer per collision, is ``conveniently taken''~\cite{BDMPSZ} as the Debye screening mass acquired by gluons in the medium. In this, the original BDMPSZ formalism~\cite{BDMPSZ}, the energy loss of an outgoing parton, $-dE/dx$,  
per unit length ($x$) of a medium with total length $L$, due to coherent gluon bremsstrahlung is proportional to the 4-momentum-square transferred to the medium and takes the form:
\begin{equation}
{-dE \over dx }\simeq \alpha_s \mean{q^2(L)}=\alpha_s\, \hat{q}\, L=\alpha_s\, 
\mu^2\, L/\lambda_{\rm mfp} \qquad , \label{eq:dEdx}
\end {equation}
so that the total energy loss in the medium goes like $L^2$~\cite{BDMPS2}. 

Additionally the accumulated transverse momentum-square, $\mean{k_{T}^2}$, for a gluon traversing a length $L$ in the medium is well approximated by $\mean{k_{T}^2}\approx\mean{q^2(L)}=\hat{q}\, L$.  This leads to a remarkable relationship~\cite{BDMPSZ} between the energy loss and di-jet broadening (``acoplanarity~\cite{GyulassyPlumerPLB243}''):
\begin{equation}
{-dE \over dx }\simeq \alpha_s \mean{k_{T}^2} \qquad , \label{eq:dEdxkT2}
\end{equation}
which is thought to be independent of the dynamics of the individual scatterings in p\QCD\ and thus should be expected to hold equally in a finite length \QGP\ and CNM~\cite{BDMPS3}. 
A simple estimate shows that the $\mean{k_{T}^2}\approx\hat{q}\,L$ should be observable at RHIC via the broadening of di-hadron azimuthal correlations.  Assume that for a trigger particle with $p_{T_t}$ the away-parton traverses slightly more than half the  14 fm diameter medium for central collisions of Au$+$Au, say 8 fm. With a $\hat{q}=1$ GeV$^2$/fm~\cite{JETcollabPRC90}, this would correspond to $\mean{k_{T}^2}=\hat{q}\,L=8$ (GeV/c)$^2$, compared to the measured~\cite{ppg029} $\mean{k_T^2}=8.0\pm 0.2$ (GeV/c)$^2$ for di-hadrons in $p+p$  collisions\footnote{In both cases the azimuthal projection is only half the $\mean{k_{T}^2}$ in $p+p$ or from $\hat{q}$.}  with roughly the same $p_{T_t}$ and $p_T^{\rm assoc}$. This should be visible as a width of the $p_T^{\rm assoc}$ azimuthal distribution $\sim\sqrt{2}$ larger in Au$+$Au   than in $p+p$  collisions at \sqsn=200 GeV. 

However, there is no direct evidence as yet for broadening of di-hadron or di-jet correlations from the effect of $\hat{q}$ in either $d+$Au ~\cite{ppg039} or Au$+$Au   collisions at RHIC, where the principal difficulty in Au$+$Au   stems from the systematic uncertanties due to the collective flow background of the medium, $v_2,v_3\ldots v_n$ for di-hadron measurements;  nor at LHC, where the very large jet $\pT\simeq 100$ GeV/c, for di-jet measurements, may have obscured this signal. 
\section{Discovery of the real \QGP\ jet quenching, RHIC's main claim to fame.} 

\begin{figure}[!h]
\vspace*{-0.3pc}        \centering
\includegraphics[height=0.26\textheight]{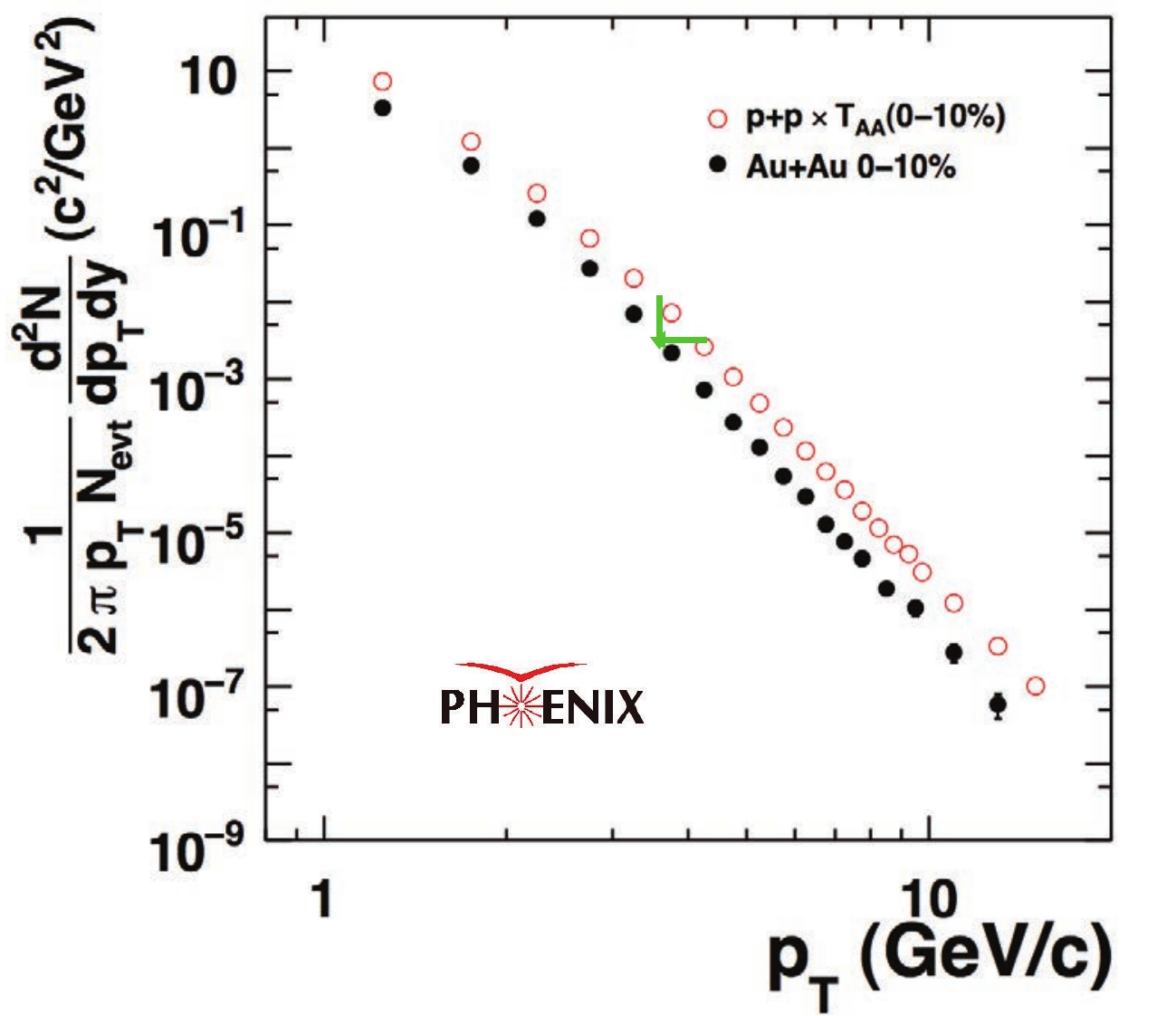}
\hspace*{-0.011\textwidth} \includegraphics[height=0.26\textheight]{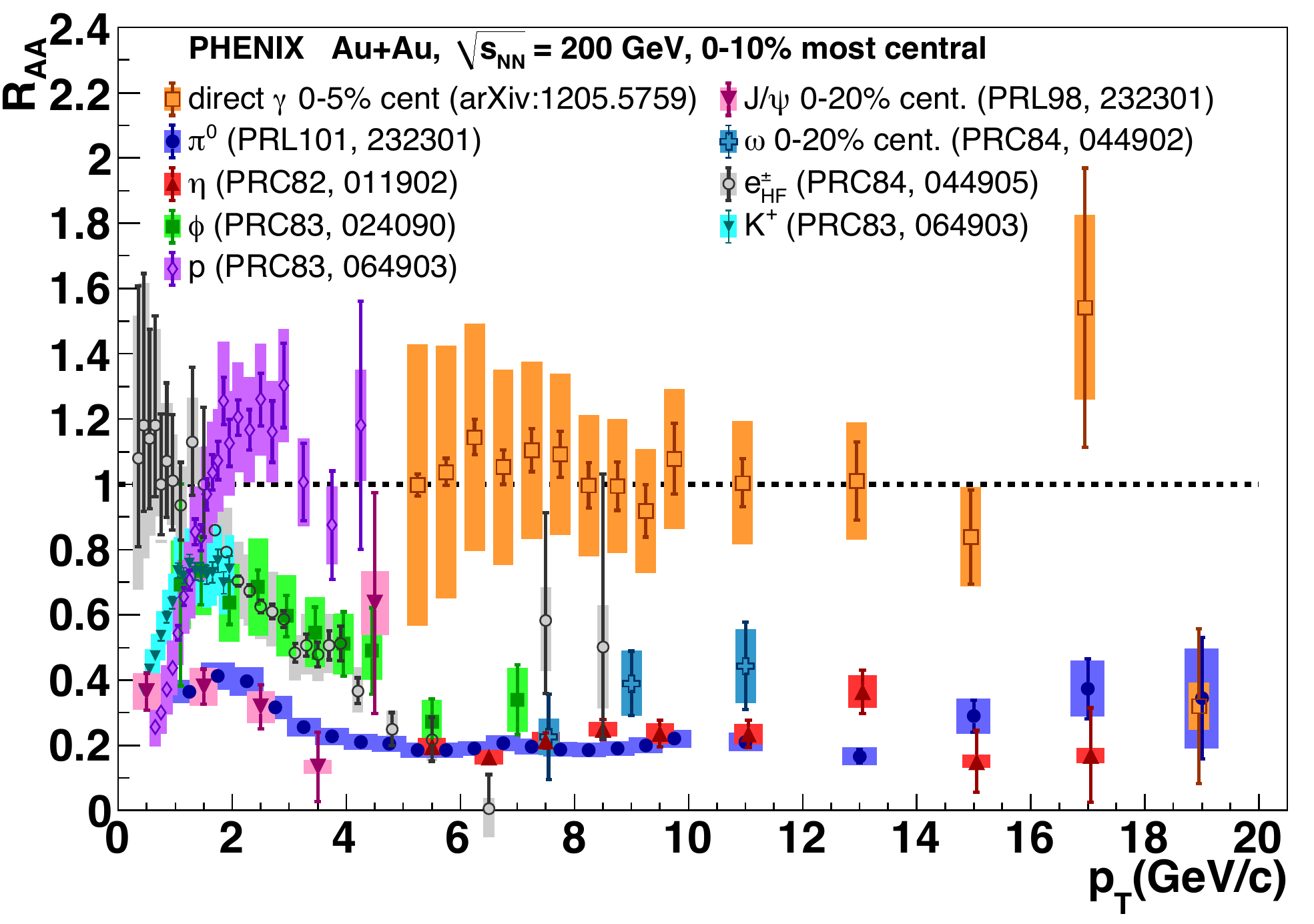}\vspace*{-0.5pc}
\caption{  a) (left) Log-log plot of invariant yield of $\pi^0$ at $\sqrt{s_{NN}}=200$ GeV as a function of transverse momentum $p_T$ in $p$$+$$p$ collisions, multiplied by $\mean{T_{\rm{AA}}}$ for Au$+$Au   central (0--10\%) collisions, compared to the Au$+$Au   measurement~\cite{ppg054}. Vertical arrow is for $R_{\rm{AA}}(p_T)$, horizontal arrow for $\delta p_T'$. b) (right) $R_{\rm{AA}}(p_T)$ for all identified particles so far measured by PHENIX in Au$+$Au   central collisions at $\sqrt{s_{NN}}=200$ GeV.}
\label{fig:Tshirt}
\end{figure}

   The discovery at RHIC~\cite{ppg003} that $\pi^0$'s produced at large transverse momenta are suppressed in central Au$+$Au   collisions by a factor of $\sim5$ compared to pointlike scaling from $p$$+$$p$ collisions is arguably {\em the}  major discovery in Relativistic Heavy Ion Physics. For $\pi^0$ (Fig.~\ref{fig:Tshirt}a)~\cite{ppg054} the hard-scattering in $p$$+$$p$ collisions is indicated by the power law behavior $p_T^{-n}$ for the invariant cross section, $E d^3\sigma/dp^3$, with $n=8.1\pm 0.1$ for $p_T\geq 3$ GeV/c.  The Au$+$Au   data at a given $p_T$ can be characterized either as shifted lower in \pT by $\delta p_T'$ from the pointlike scaled $p$$+$$p$ data at $p'_T=p_T+\delta p_T'$, or shifted down in magnitude, i.e. suppressed. In Fig.~\ref{fig:Tshirt}b, the suppression of the many identified particles measured by PHENIX at RHIC is presented as the Nuclear Modification Factor, 
$R_{\rm{AA}}(p_T)$, the ratio of the yield of e.g. $\pi$ per central Au$+$Au   collision (upper 10\%-ile of observed multiplicity)  to the pointlike-scaled $p$$+$$p$ cross section at the same $p_T$, where $\mean{T_{\rm{AA}}}$ is the average overlap integral of the nuclear thickness functions: 
   \begin{equation}
  R_{\rm{AA}}(p_T)=\frac{(1/N_{\rm{AA}})\;{d^2N^{\pi}_{\rm{AA}}/dp_T dy}} { \mean{T_{\rm{AA}}}\;\, d^2\sigma^{\pi}_{\rm{pp}}/dp_T dy} \quad . 
  \label{eq:RAA}
  \end{equation}

The striking differences and similarities of $R_{\rm{AA}}(p_T)$ in central Au$+$Au   collisions for the many particles measured by PHENIX  (Fig.~\ref{fig:Tshirt}b) illustrate the importance of particle identification for understanding the physics of the medium produced at RHIC. Notable are that ALL particles are suppressed for $\pT>4$ GeV/c (except for direct-$\gamma$ which are not coupled to color), even electrons from $c$ and $b$ quark decay; with one notable exception: the protons are enhanced for $2\leq \pT\leq 4$ GeV/c, called the baryon anomaly, although recently the same Cronin-like effect has been seen in $d+$Au  collisions~\cite{ppg146}.
\subsection{$\mathbf{\delta p_T'/p_T'}$, the fractional shift in the $\mathbf{p_T'}$ spectrum}
\begin{figure}[!h]
   \begin{center}
\includegraphics[width=0.49\textwidth]{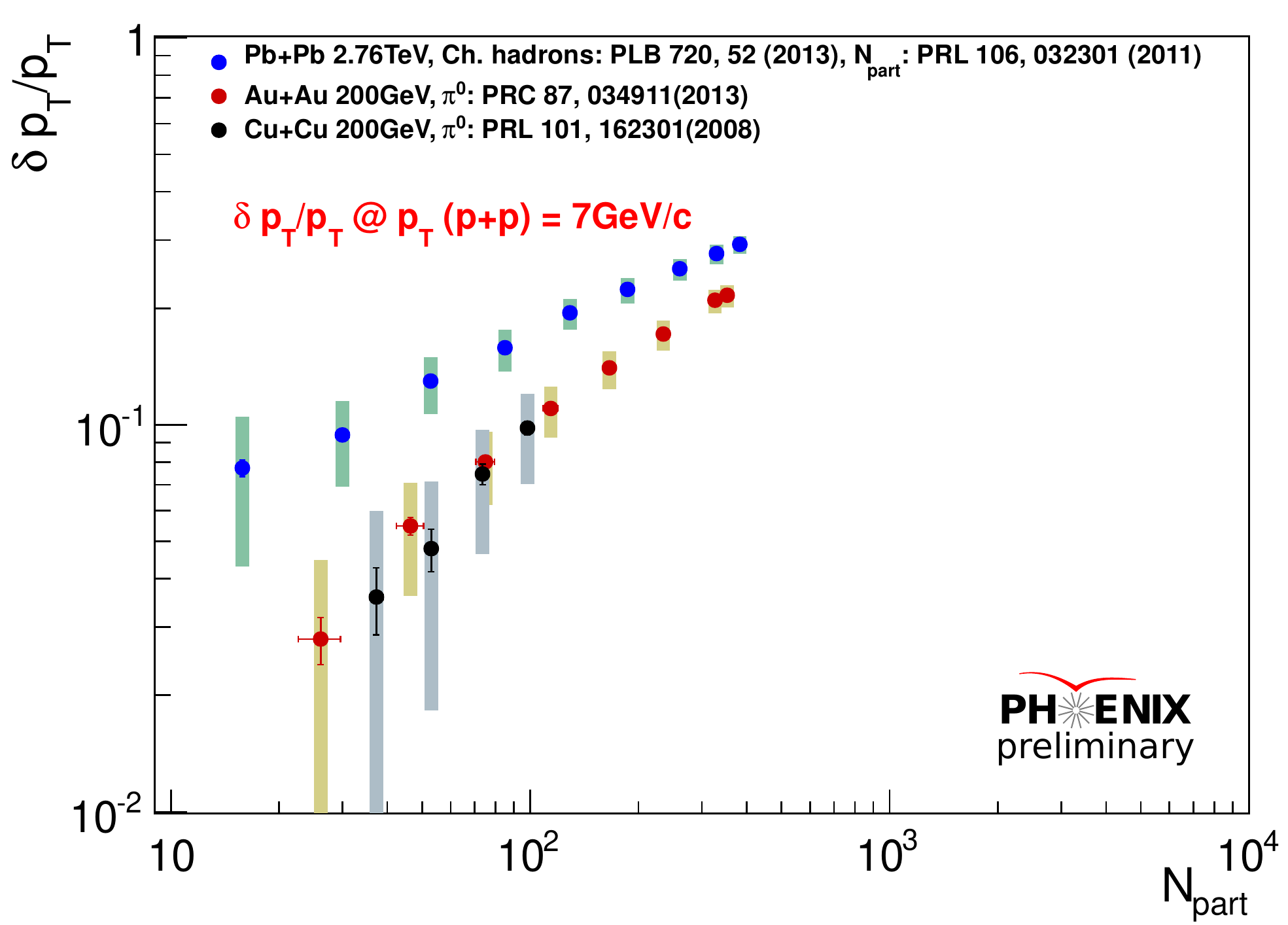}
\includegraphics[width=0.49\textwidth]{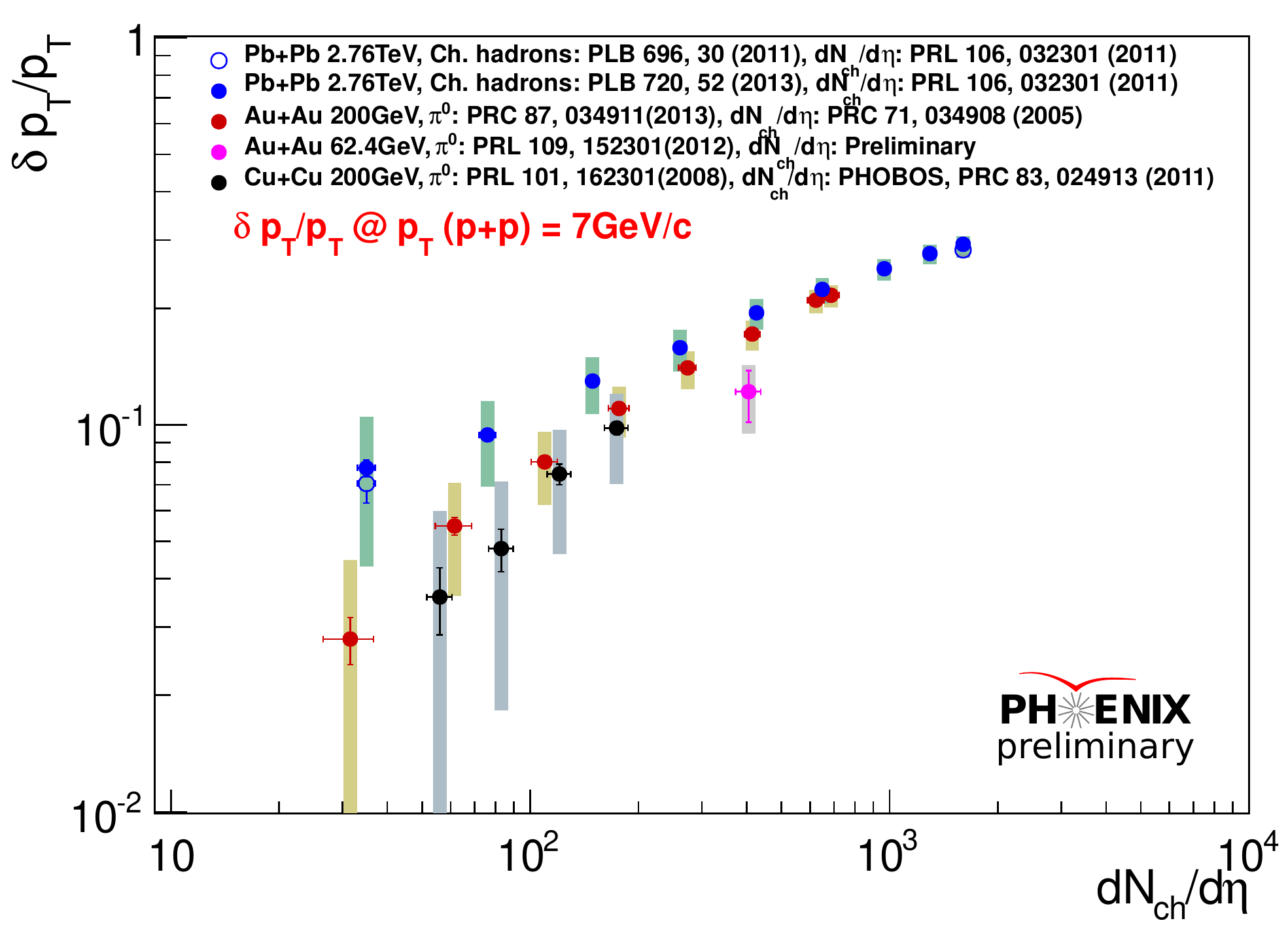}
\end{center}\vspace*{-1.0pc}
\caption[]{  Plots from PHENIX~\cite{TakaoNantes} of $\delta p_T'/p_T'$ at $p_T'\equiv p_T(p+p)=7$ GeV for $\pi^0$ (RHIC) and charged hadrons (LHC): a) as a function of centrality (\Npart), b) as a function of $d\Nch/d\eta$.  
\label{fig:dpTpT}}\vspace*{-0.1pc}
\end{figure}

After more than a decade of using the ratio $R_{\rm{AA}}$, we are now paying more attention to $\delta p_T'/p_T'$, the fractional shift of the $p_T'$ spectrum, as an indicator of energy loss in the \QGP\ Fig.~\ref{fig:dpTpT}~\cite{TakaoNantes}. 
For a constant fractional energy loss, which is true at RHIC in the range $6<p_T<12$ GeV/c (as shown in Fig.~\ref{fig:Tshirt}a where the $p$$+$$p$ reference and Au$+$Au measurement are parallel on a log-log plot) there is a simple relationship between $R_{\rm{AA}}$, $\delta p_T'/p_T'$ and $n$,  the power in the invariant $p_T$ spectra:
\begin{equation}
R_{\rm{AA}}(p_T')=R_{\rm{AA}}(p_T)=(1-\delta p_T'/p_T')^{n-2}\qquad . \label{eq:RAAdelta}
\end{equation}
Using  $\delta p_T'/p_T'$ is important for comparison to the LHC measurements where the power is $n\approx 6$ compared to $n=8.1$ at RHIC, so that the same $R_{\rm{AA}}$ does not mean the same $\delta p_T'/p_T'$. Strictly $\delta p_T'/p_T'$ is not a measure of the parton energy loss in the \QGP\ but is used as a proxy.   
Figure \ref{fig:dpTpT}a shows that $\delta p_T'/p_T'$ at $p_T'=7$ GeV/c for RHIC and LHC both increase monotonically with centrality (\Npart) but is a factor of 2 to 1.4 larger at LHC, depending on centrality, a likely indication of a hotter and/or denser medium. Figure \ref{fig:dpTpT}b attempts to determine whether $\delta p_T'/p_T'$ is a universal function of the charged particle density, $d\Nch/d\eta$ at both RHIC and LHC, as suggested by Edward Shuryak at this meeting last year~\cite{ShuryaktoMJT2013}. The dependence is not quite universal. A fit of $\delta p_T/p_T\propto (d\Nch/d\eta)^\alpha$ gives $\alpha\approx 0.35$ at LHC and 0.55 at RHIC, although the data at $\sqsn=200$ GeV and 2.76 TeV do appear to merge for $(d\Nch/d\eta)\geq 300$. Hopefully, measurements of $\delta p_T/p_T$ will eventually lead to the determination of $dE/dx$ of partons in the \QGP. 
\section{STAR jet and jet-hadron measurements}
\subsection{Jet-hadron correlations as a proxy for di-jet broadening}
\begin{figure}[!h] 
      \centering
\raisebox{0.0pc}{\includegraphics[width=0.45\linewidth]{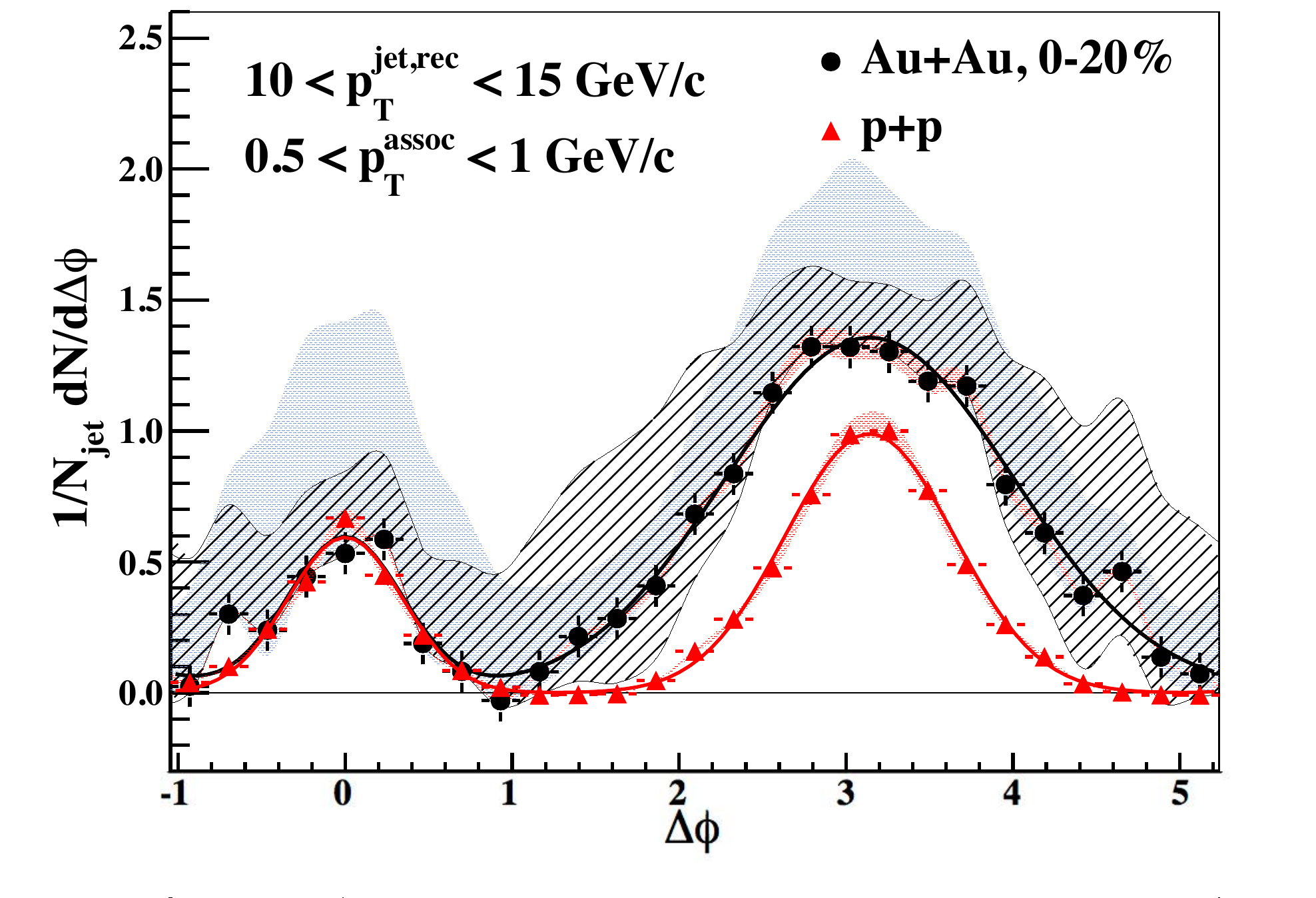}} 
\raisebox{0.2pc}{\begin{minipage}[b]{0.44\linewidth}
\includegraphics[width=\linewidth]{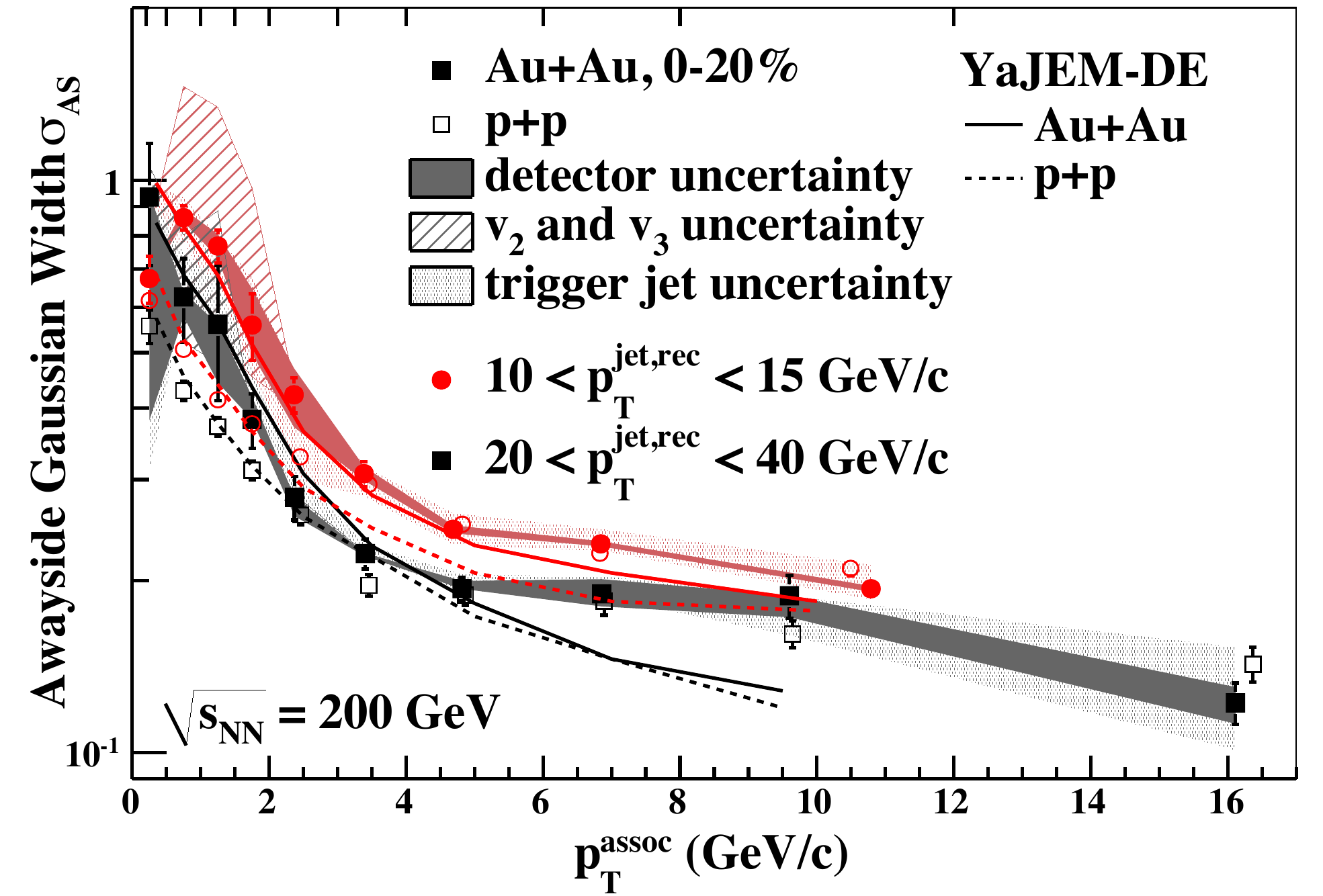} \end{minipage}}
\caption[] {a) (left) Azimuthal correlation $(1/N_{\rm jet} dN/d\Delta\phi)$ in Au$+$Au and $p+p$ with systematic uncertainties shown~\cite{STARJet-h-PRL112}. b)(right) Awayside rms width, $\sigma_{\rm AS}$, as a function of $p^{\rm assoc}_T$~\cite{STARJet-h-PRL112}.} 
      \label{fig:Caines}
   \end{figure}

Admittedly, measuring jets at RHIC at \sqsn=200 GeV is much harder than at LHC at \sqsn=2.76 TeV: the cross section in the relevant region is $\gsim 100$ times larger at LHC while the soft physics background is only a factor of 2 larger~\cite{ALICEPRL106}. Nevertheless, the principal difficulty in observing the broadening of di-jet or di-hadron azimuthal correlations by the transport coefficient $\hat{q}$ of the \QGP\ stems from artifacts with names such as ``Mach Cone'', ``Ridge'', ``Head and Shoulders'' which are now known to be due to the modulation of the soft physics background by collective flow with both even and odd harmonics~\cite{AlverRoland}.
Of course, understanding that the extra ``bumps'' in the correlation function are due to odd harmonics still requires one to know the values of these harmonics in order to subtract them. This is still the largest systematic uncertainty in attempts to observe the $\hat{q}$-broadening, for instance, the most recent attempt by STAR using jet-hadron correlations~\cite{STARJet-h-PRL112} (Fig.~\ref{fig:Caines}).   
When the full systematic uncertainties, including those on $v_2$ and $v_3$ (Fig.~\ref{fig:Caines}a), are taken into account,  the result for the medium induced broadening of the away-side widths, $\sigma_{\rm AS}$, in Au$+$Au relative to $p+p$ (Fig.~\ref{fig:Caines}b) which looked significant in the preliminary results, as shown last year~\cite{MJTHPT2013proc}, become only ``suggestive of medium-induced broadening~\cite{STARJet-h-PRL112}'' in the final result  because ``they are highly dependent on the shape of the subtracted background~\cite{STARJet-h-PRL112}'', notably the $v_2$ and $v_3$ of the trigger jets. 

\subsection{At last: jet measurements in Au$+$Au at RHIC in 2014?} 
\begin{figure}[!htb] 
\begin{center}
 \raisebox{0.0pc}{\includegraphics[width=0.8\textwidth]{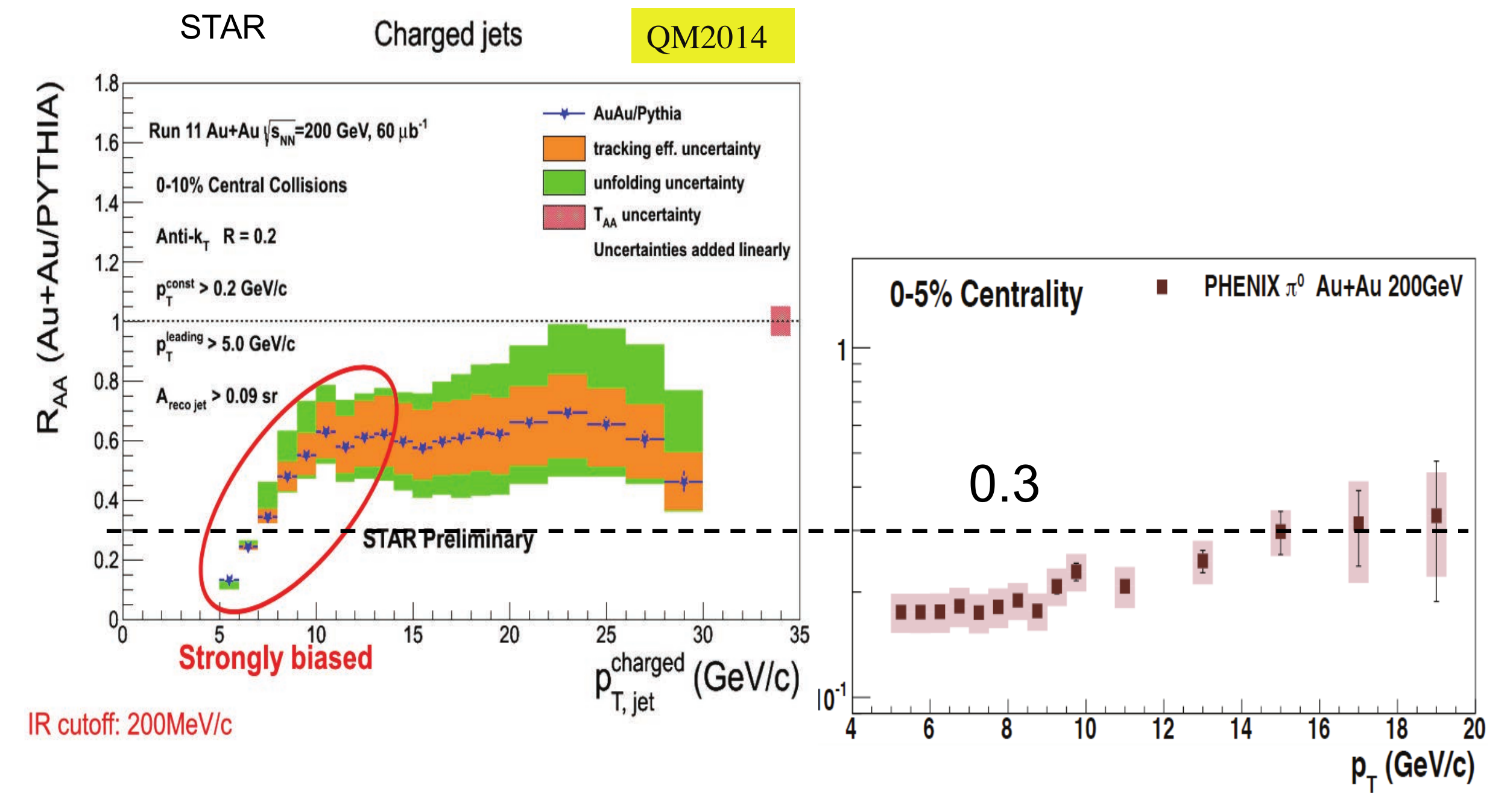}}
\end{center}\vspace*{-1.3pc}
\caption[]{  a) (left) STAR $R_{\rm{AA}}$ for charged jets at \sqsn=200 GeV in central Au$+$Au collisions (see details in legend) compared to b) $R_{\rm{AA}}$ for PHENIX $\pi^0$. The dashed line at 0.3 is the maximum $R_{\rm{AA}}$ for $\pi^0$ in this $p_T$ range.\label{fig:STARjetvssingle}}\vspace*{-1.0pc}
\end{figure}

Some interesting new jet measurements in Au$+$Au collisions at RHIC were presented at Quark Matter 2014 in a plenary review talk on jets by Yen-Jie Lee~\cite{YJLeeQM2014} who works on CMS. Figure~\ref{fig:STARjetvssingle} shows that the STAR charged jets in a cone with $R=0.2$ have much less suppression ($R_{\rm{AA}}\gg 0.3$) than $\pi^0$ ($0.2\leq R_{\rm{AA}}\leq 0.3$) in the range $10< p_T<20$ GeV. \vspace*{-0.7pc}

\begin{figure}[!h] 
\begin{center}
 \raisebox{0.0pc}{\includegraphics[width=0.80\textwidth]{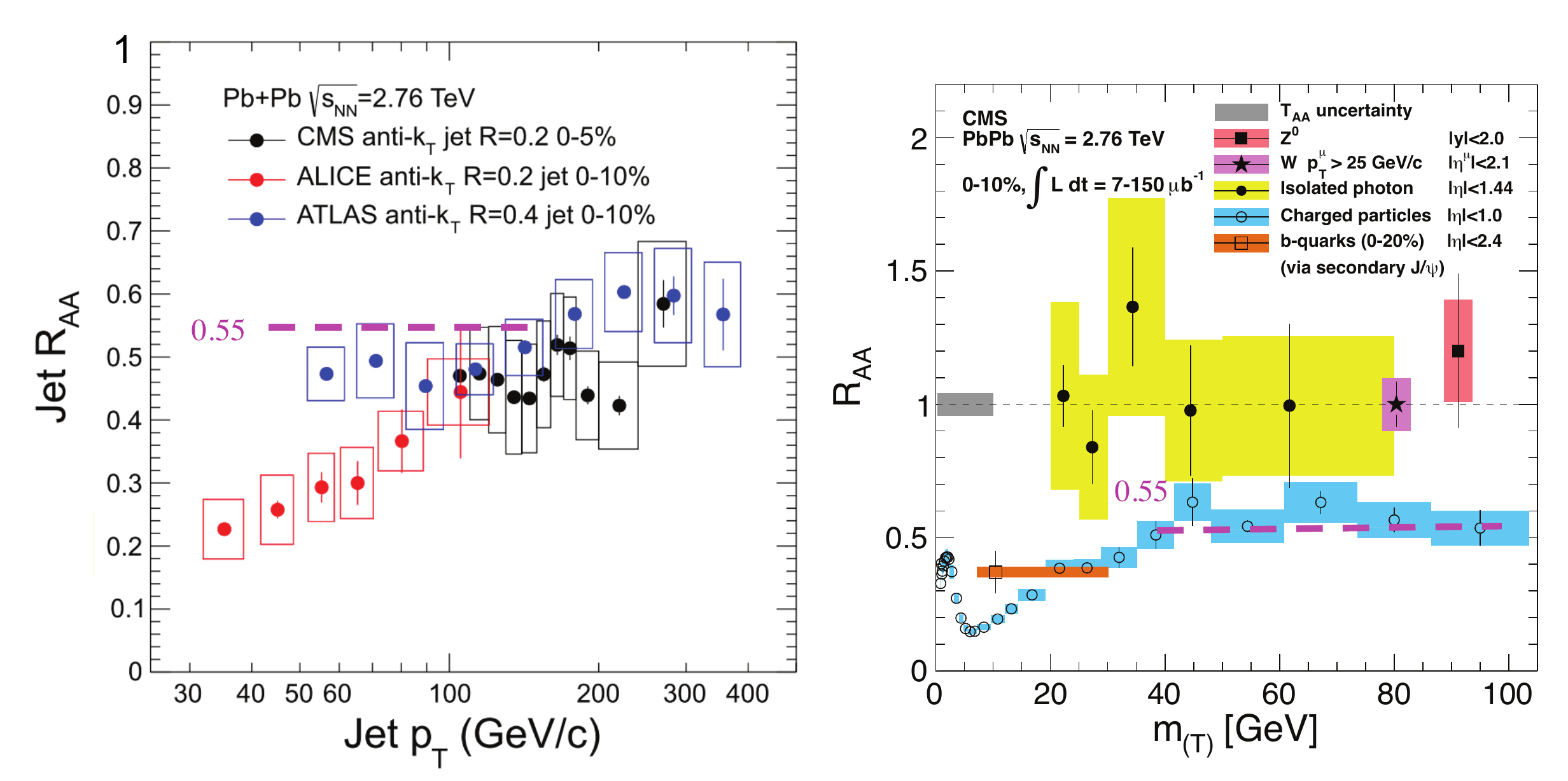}}
\end{center}\vspace*{-1.0pc}
\caption[]{  a) (left) $R_{\rm{AA}}$ for jets at \sqsn=2.76 GeV by CMS and ALICE  compared to b) CMS $R_{\rm{AA}}$ for charged hadrons ($R_{\rm{AA}}\approx 0.55$), $b$-quarks and 3 favorite Electro-Weak Bosons~\cite{YJLeeQM2014}.\label{fig:CMSjetvssingle}}\vspace*{-2.0pc}
\end{figure}

This is quite different from jets at the LHC (Fig.~\ref{fig:CMSjetvssingle}) which have comparable or smaller $R_{\rm{AA}}$ than charged particles from jet fragmentation in the range $30<p_T<100$ GeV.  Note that the $\gamma$, $W$ and $Z^0$ bosons in Fig.~\ref{fig:CMSjetvssingle}b which are not coupled to color are not suppressed.

\begin{figure}[!htb]\vspace*{-0.8pc} 
\begin{center}
 \raisebox{0.0pc}{\includegraphics[width=0.9\textwidth]{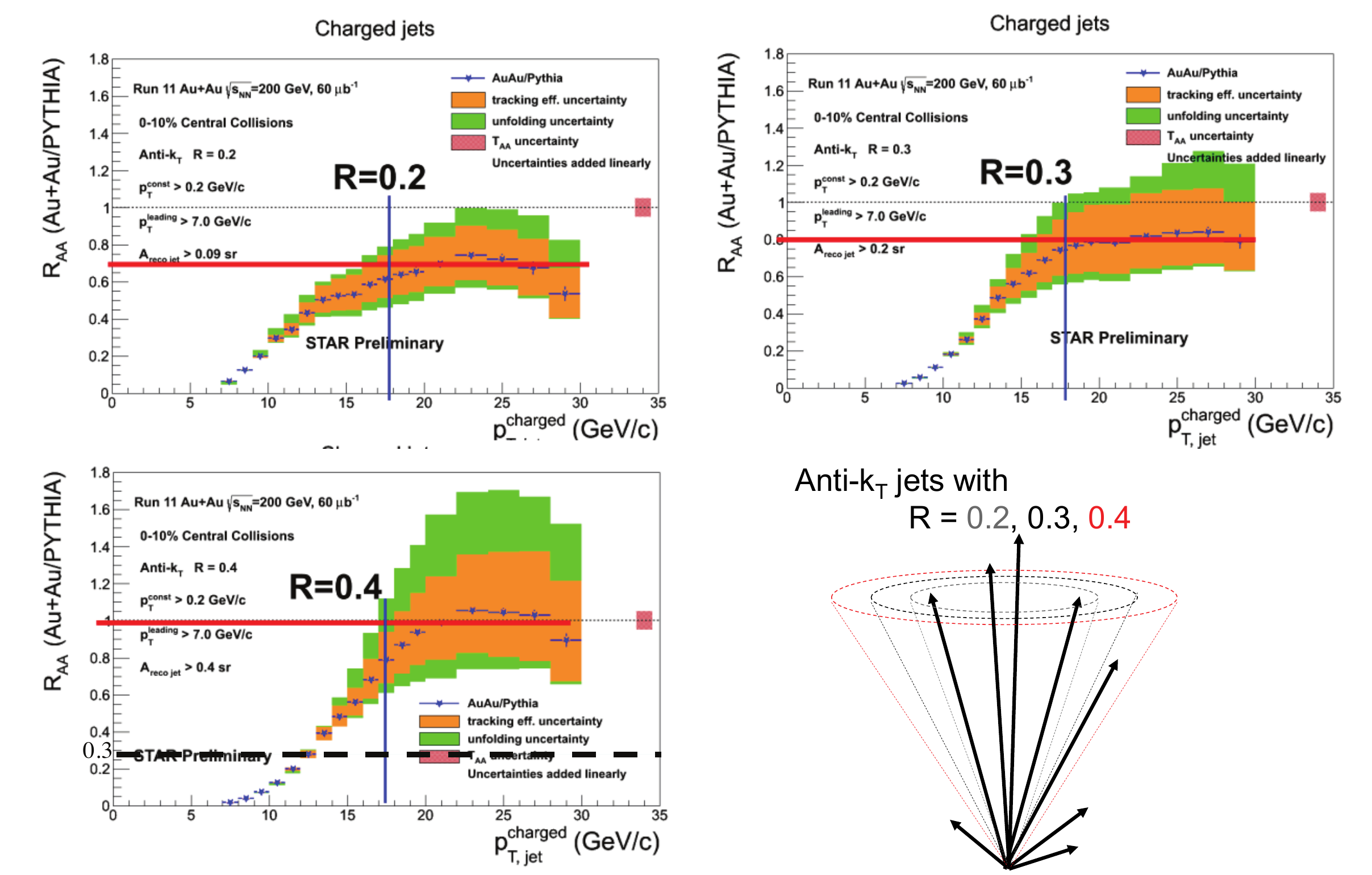}}
\end{center}\vspace*{-1.5pc}
\caption[]{  STAR $R_{\rm{AA}}$ for charged jets at \sqsn=200 GeV in central Au$+$Au collisions for 3 different jet cones with $R=0.2,0.3,0.4$ (see details in legend and sketch)~\cite{YJLeeQM2014}. \label{fig:STAR3Rs}}\vspace*{-0.5pc}
\end{figure}

For STAR, the disagreement of the jet and single particle $R_{\rm{AA}}$ gets worse as the jet cone is increased from $R$=0.2 to 0.3 to 0.4 (Fig.~\ref{fig:STAR3Rs}). Some people would say that this is great because all the jet fragments and/or any energy lost in the \QGP\ by the originating parton have been captured in the $R$=0.4 cone. Skeptics like myself can hardly wait to see what happens when the jet cone is further increased. After 14 runs at RHIC, the jet learning curve in Au$+$Au central collisions still has a way to go.

The good news for the future is that a new detector, now called sPHENIX, to find jets by the more traditional method using hadron calorimetry has been proposed, is moving along on the approval process and is on the schedule at RHIC for partial commisioning in 2019.
\section{Kari Eskola once asked me whether I believed in \QCD}
In the 4th meeting in this series, in Prague in 2009, Kari Eskola asked me whether I believed in \QCD\ after I expressed doubt about some calculation. I answered, ``Of course I believe in \QCD; but I am skeptical of many calculations that claim to be \QCD.'' Such calculations are still being made which I learned about by reading Jan Rak's talk at a recent conference~\cite{JanCrete2014}. 
\subsection{Another wrong calculation claiming to be \QCD}
\begin{figure}[!t] 
      \centering
\raisebox{0.3pc}{\includegraphics[width=0.40\linewidth, height=0.50\textwidth]{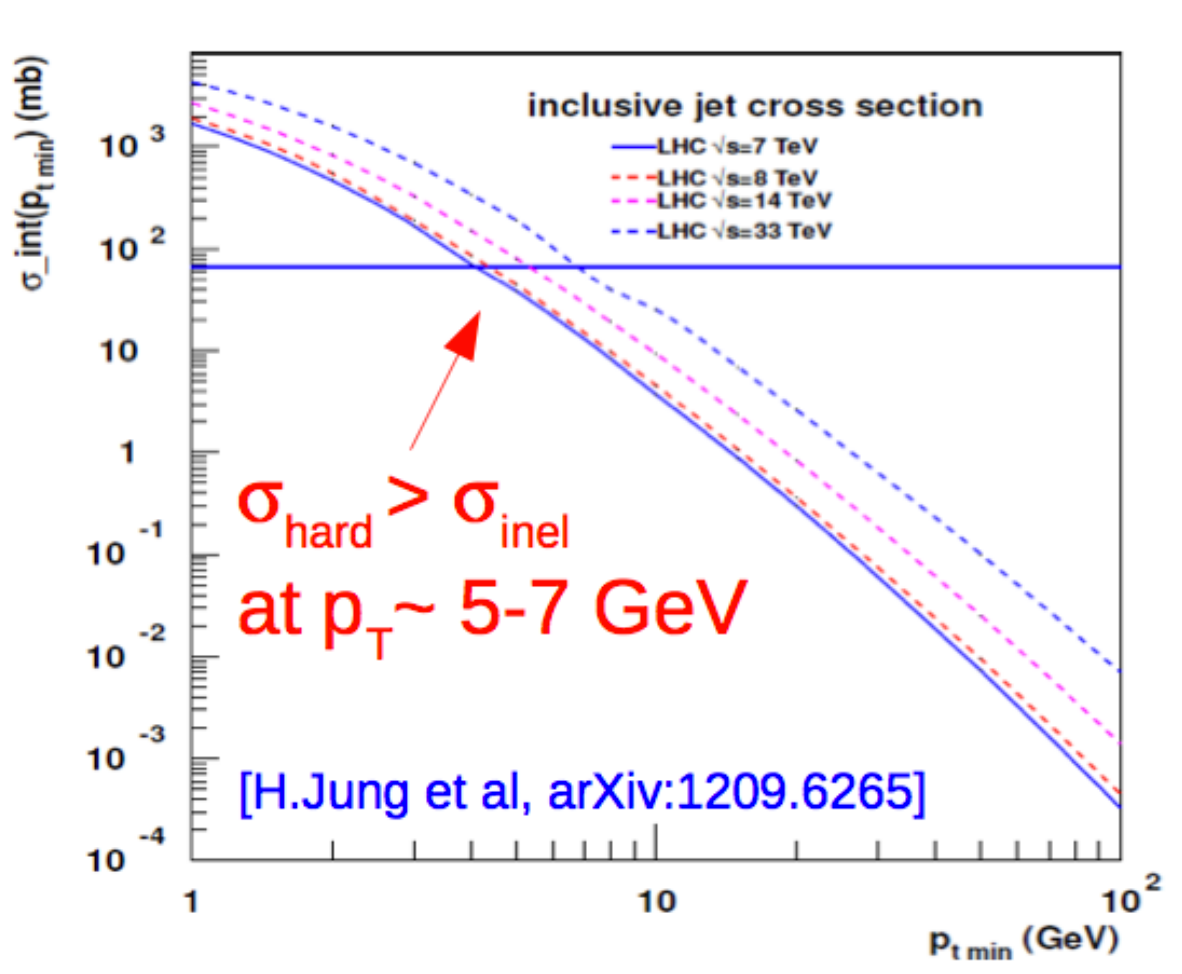}}\hspace{1pc}
\raisebox{0.0pc}{\includegraphics[width=0.40\linewidth]{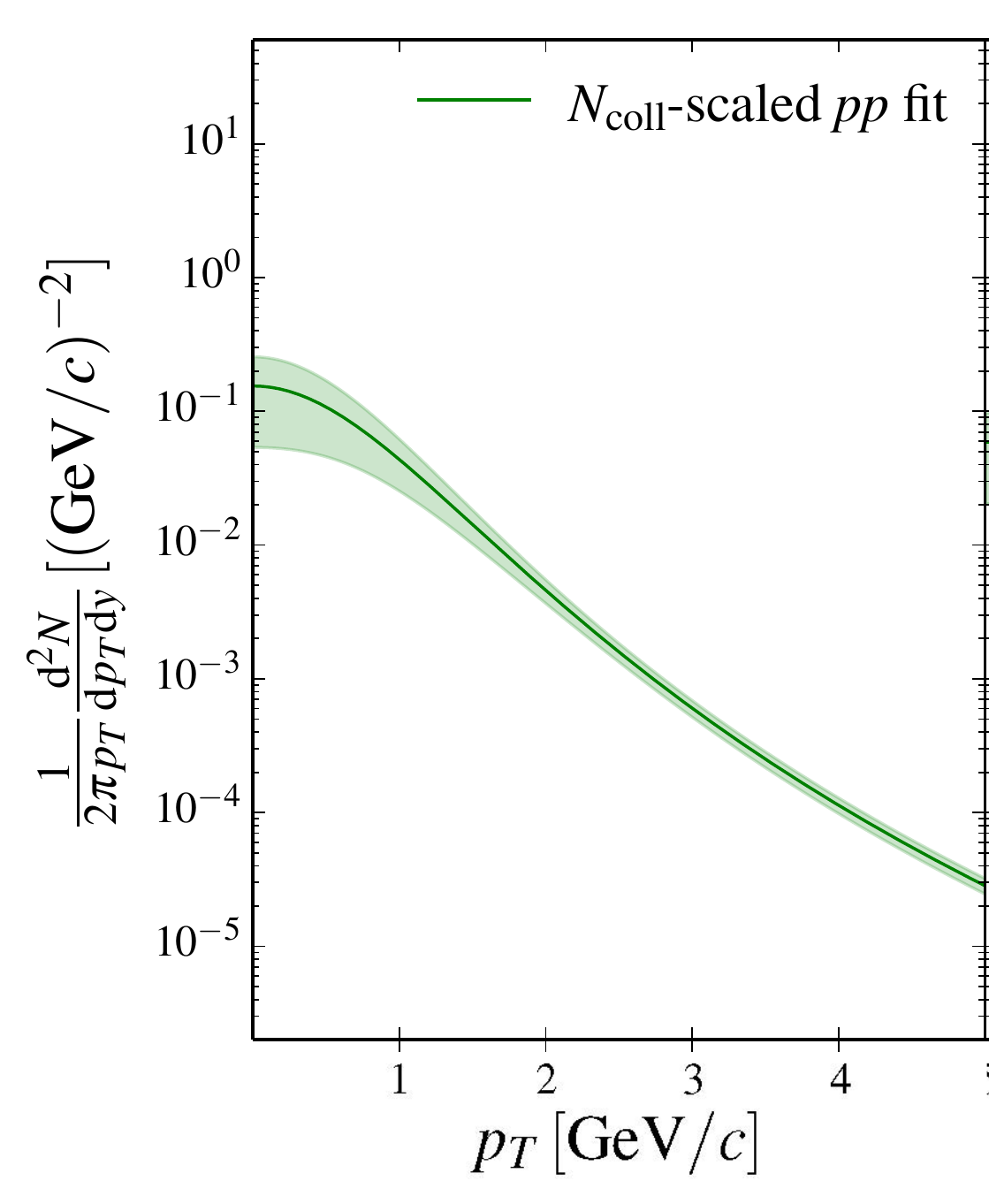}}
\caption[] {a) (left) Integrated inclusive jet cross section for $\pT\geq p_{T_{\rm min}}$ as a function of $p_{T_{\rm min}}$ from Ref.~\cite{JanJung} for several \sqsn indicated. b) (right) $(Ed^3\sigma/dp^3)/(0.054$ mb) for direct-$\gamma$ production in $p+p$ collisions at \sqs=200 GeV. The distribution is scaled for comparison to Au$+$Au central (0-20\%) measurements~\cite{PXppg162}.  }
      \label{fig:JanJung}
   \end{figure}

Figure~\ref{fig:JanJung}a~\cite{JanJung} shows a supposed \QCD\ calculation of the inclusive jet cross section in $p+p$ collisions at \sqsn=7--33 TeV in which the integrated inclusive jet cross section exceeds the inelastic cross section. This is normal for inclusive measurements, e.g. single particle spectra, where the integral of the inclusive cross section equals the interaction cross section times the mean multiplicity, but is well known not to happen in hard-scattering.   Nature (i.e. non-perturbative \QCD) finds a way to stop the $p_T^{-n}$ divergence, which flattens for $\pT\lsim 3$ GeV/c as shown for direct-$\gamma$ production in Fig~\ref{fig:JanJung}b. The same flattening happens for the \pT distribution of Drell-Yan lepton pair production~\cite{CFSPRD23}.
Even though the authors of Pythia provided ``a phenomenological modification of the low-$p_T$ behiavior of the jet cross section'' to agree with the actual \QCD\ behavior, the Pythia `calculators'~\cite{JanJung} decided not to use it and got a ridiculous answer, once again confirming my response to Kari. 
   
\subsection{Direct-$\gamma$ production, real \QCD\ calculations and $x_T$ scaling}   
\begin{figure}[!t] 
      \centering
\raisebox{1.0pc}{\includegraphics[width=0.45\linewidth]{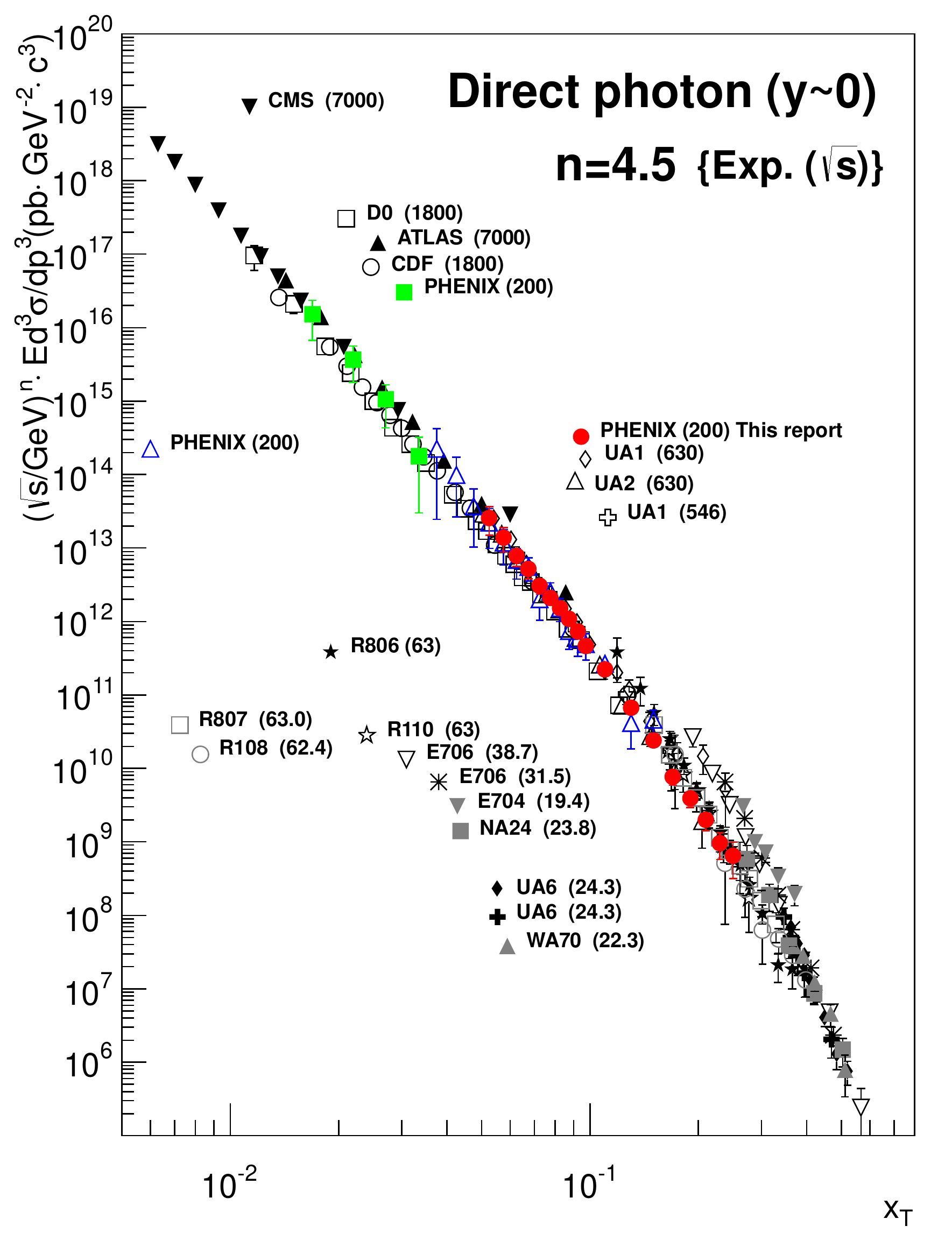}}\hspace{1pc}
\raisebox{0.0pc}{\includegraphics[width=0.51\linewidth]{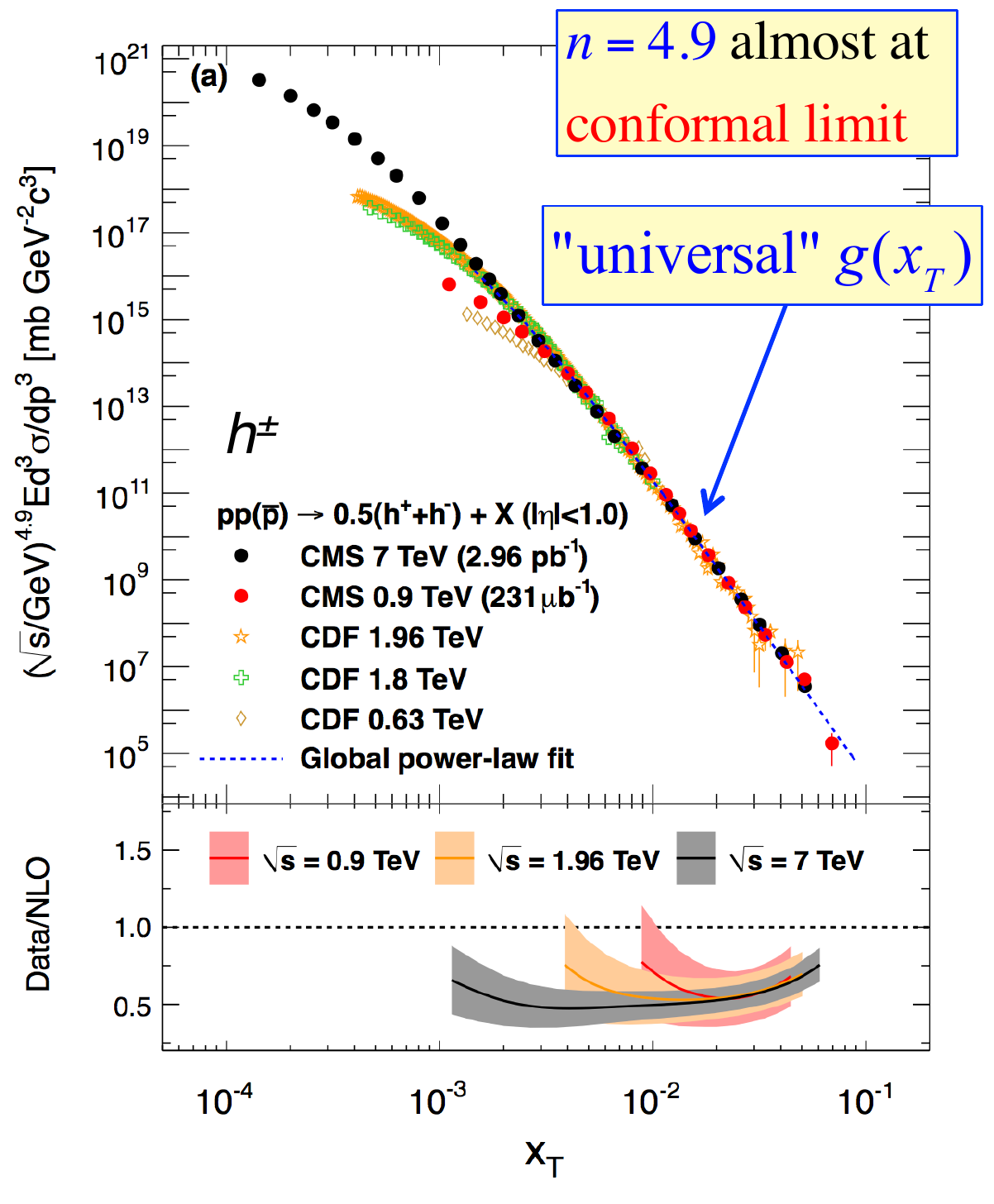}}\hspace{1pc}
\caption[] {a) (left) Direct-$\gamma$ measurements plotted as $\sqrt{s}^{\, n_{\rm eff}} \times Ed^3\sigma/dp^3$ at $x_T
\equiv 2\pT/\sqs$ with $n_{\rm eff}=4.5$~\cite{PXdirgppPRD86}. The legend gives the experiment and $\sqrt{s}$. b)(right) $x_T$ scaling for inclusive charged particles at $\sqs\ \gsim\ 1$ TeV with $n_{\rm eff}=4.9$~\cite{CMSJHEP2011}. }\vspace*{-1.0pc}
      \label{fig:alldirgxT}
   \end{figure}

My favorite \QCD\ reaction is direct-$\gamma$ production via the subprocess $g+q\rightarrow \gamma+q$. This is much better than jet production to test \QCD\ calculations as well as to measure  parton energy loss in the \QGP\ for several reasons: i) the $\gamma$ participates directly in the hard-scattering and then emerges freely and unbiased from the reaction, with no accompanying particles, and passes unaffected through the medium to a detector where its energy can be measured precisely; ii) the transverse momentum of the jet from the outgoing quark at the reaction point is equal and opposite to that of the $\gamma$, thus is also precisely known (modulo $k_T$);
iii) for LO p\QCD\ calculations of the direct-$\gamma$ inclusive spectrum, no fragmentaton functions are needed---a major advantage over jet and single particle calculations. This is illustrated in Fig.~\ref{fig:alldirgxT} where $x_T$ scaling is presented for both inclusive direct-$\gamma$ (Fig.~\ref{fig:alldirgxT}a) over a large range of \sqs in $p+p$ and $\bar{p}+p$ collisions and for inclusive charged particles at 3 values of \sqs (Fig.~\ref{fig:alldirgxT}b).  

$x_T$ scaling~\cite{BBGPLB42,CahalanPRD11} provides a totally data driven test of whether p\QCD\ or some other underlying subprocess is at work, without the need to know the details of the structure functions, fragmentation function and coupling constant, as well as providing a compact quantitative way to describe the data using the effective index, $n_{\rm eff}(x_T,\sqrt{s})$. The invariant cross section for inclusive single particle production can be written as:
   \begin{equation}
  E \frac{d^3\sigma}{dp^3}=\frac{d^3\sigma}{p_T dp_T dy d\phi}={1 \over {p_T^{{n_{\rm eff}(x_T,\sqrt{s})}} }  } 
F(x_T)={1\over {\sqrt{s}^{{\,n_{\rm eff}(x_T,\sqrt{s})}} } } 
\: {\rm g}(x_T) \qquad ,
\label{eq:siginv+nxt}
\end{equation}
where $E d^3\sigma/dp^3=\sigma^{\rm inv}(p_T,\sqrt{s})$ is the invariant cross section for inclusive particle production with transverse momentum $p_T$ at c.m. energy $\sqrt{s}$, and $x_T=2 p_T/\sqrt{s}$. It is important to emphasize that the effective power, $n_{\rm eff}(x_T,\sqrt{s})$, is different from the power $n$ of the invariant cross section, which varies with $\sqrt{s}$ (which it must if $x_T$ scaling is to hold).
For pure vector gluon exchange, or without the evolution of $\alpha_s$ and the structure and fragmentation functions in \QCD,  $n_{\rm eff}=4$ as in Rutherford scattering. However,  due to the non-scaling in QCD~\cite{CahalanPRD11}, the measured value of $n_{\rm eff}$ depends on the $x_T$ value and the range of $\sqrt{s}$ used. 

The point of this discussion and Fig.~\ref{fig:alldirgxT} is that the direct-$\gamma$ data are very well described by \QCD\ and $x_T$ scaling, with $n_{\rm eff}=4.5$ due to the \QCD\ evolution, while the charged particle data also follow $x_T$ scaling very well, but with a  larger $n_{\rm eff}=4.9$ due to the added non-scaling of the fragmentation functions. This shows that the charged particle cros-sections follow \QCD\ even though the NLO \QCD\ calculations miss the data by a factor of 2, which~\cite{CMSJHEP2011} ``suggests that the fragmentation functions are not well tuned for LHC energies.''
\section{Problems with centrality for reactions with very large $p_T$ in $p+$A collisions} 
\begin{figure}[!h] 
      \centering
\raisebox{0.0pc}{\includegraphics[width=0.485\linewidth]{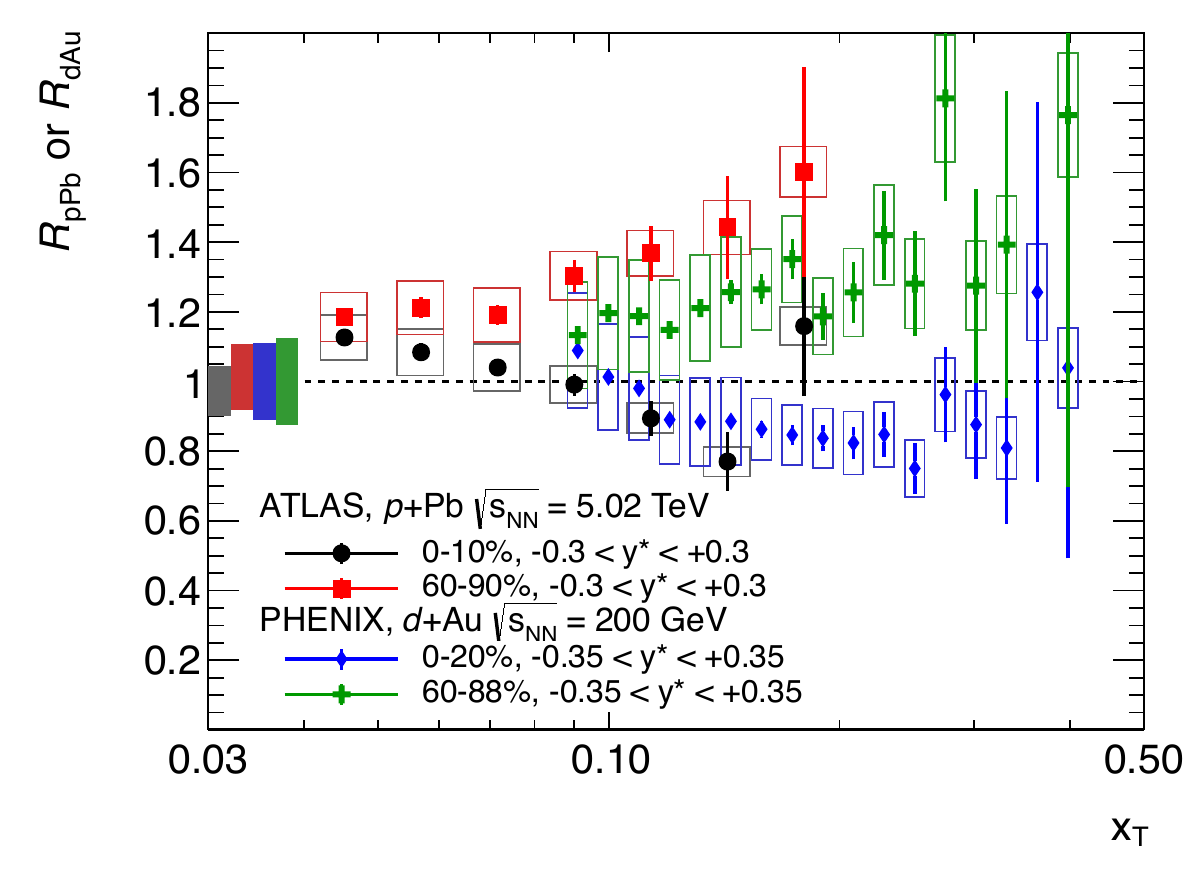}}\hspace{1pc}
\raisebox{0.0pc}{\includegraphics[width=0.48\linewidth]{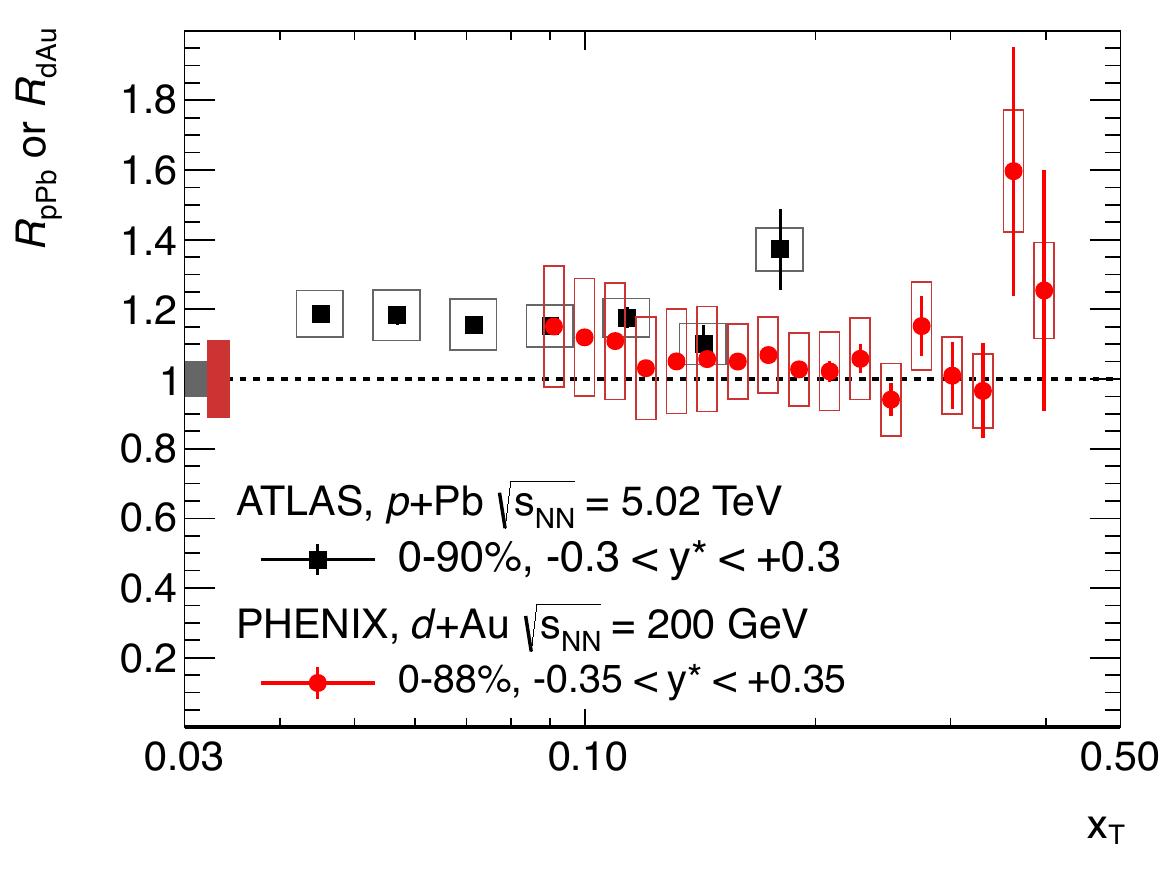}}\vspace*{-1.0pc}
\caption[] {a) (left) $R_{\rm pPb}$ (LHC) and $R_{\rm dAu}$ (RHIC) for jets~\cite{DVPthesis}, at \sqsn\ and centralities indicated. b)(right) Same for minimum bias collisions. }
      \label{fig:LHCRHICcent}\vspace*{-1.0pc}
   \end{figure}

Last year~\cite{MJTHPT2013proc}, I discussed a problem (or excitement for some people) with determining the centrality in $d+$Au at RHIC, using Beam-Beam counters at forward rapidity, $3.1<\eta<3.9$, for reactions with very large $p_T > 10$ GeV/c ($x_T>0.1$) at mid-rapidity. This year, similar methods at LHC for $p+$Pb collisions at $\sqsn=5.02$ TeV, produce a similar problem at the same $x_T$ (Fig.~\ref{fig:LHCRHICcent}a)~\cite{DVPthesis}. 
Figure ~\ref{fig:LHCRHICcent}b~\cite{DVPthesis} shows that at both LHC and RHIC, avoiding centrality cuts by using minimum bias collisions to measure $R_{\rm pA}(\pT)={\rm A}^{\alpha(p_T) -1}$ gives more reasonable results.\footnote{I use the original Cronin terminology~\cite{CroninEffect} here where the cross section for hard scattering in $p+$A collisions was represented as $\sigma_{\rm{pA}}(\pT)={\rm A}^{\alpha(p_T)} \times  \sigma_{\rm{pp}}(\pT)$, where $\alpha(\pT)\equiv 1$ for pure point-like scattering with no shadowing.} 
This is the basis for the $p+$A run at RHIC in 2015, using a few values of A to determine $\alpha(\pT)$ of minimum bias $p+$A collisions rather than make centrality cuts. 

\section{The importance of $p+p$ comparison data at the same \sqs in the same detector}
\begin{figure}[!h] 
      \centering
\raisebox{0.0pc}{\includegraphics[width=0.44\linewidth]{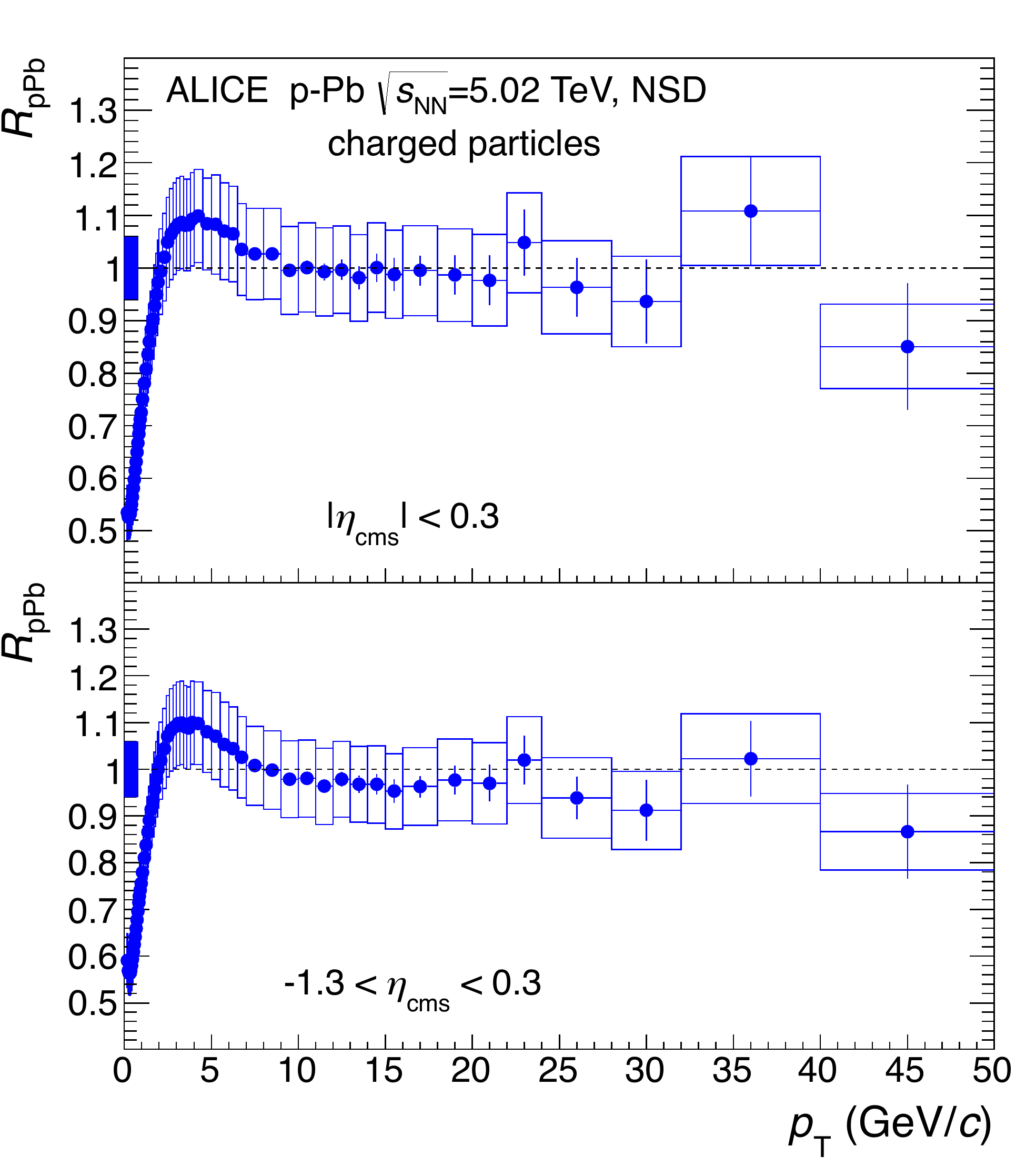}}\hspace{1pc}
\raisebox{0.0pc}{\includegraphics[width=0.52\linewidth]{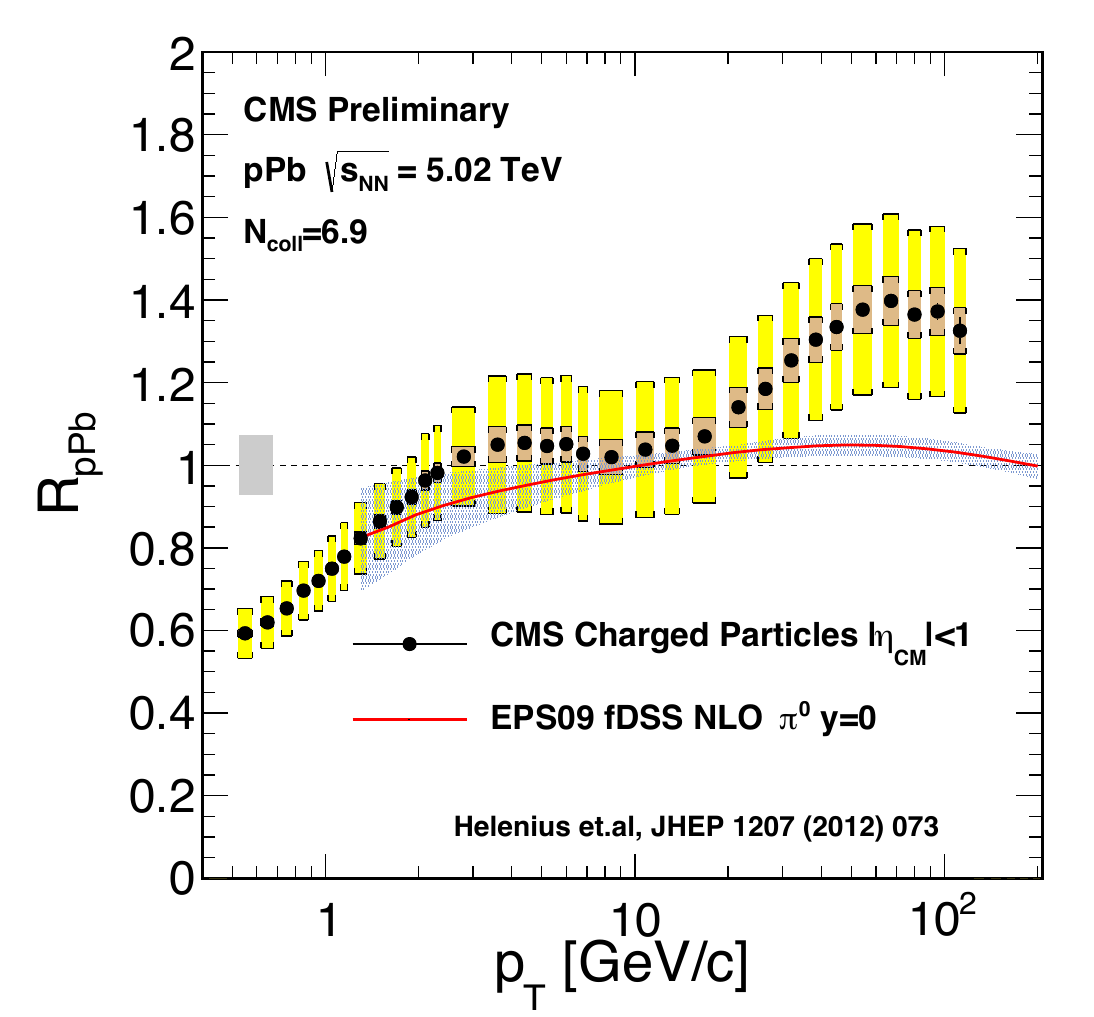}}
\caption[] {a) (left) ALICE $R_{\rm pPb}$ vs $p_T$~\cite{ALICEEPJC74} b)(right) CMS preliminary $R_{\rm pPb}$~\cite{CMSQM14}. }
      \label{fig:LHCpA}
   \end{figure}

Two LHC experiments presented measurements this year of $R_{\rm pPb}$ from $p+$Pb collisions at \sqsn=5.02 TeV from the run in 2013. It was an impressive tour-de-force for the LHC to collide particles with different Z/A in a single ring; but the price the experimenters paid was that they had no comparison $p+p$ data at the same \sqs. The results from the ALICE and CMS experiments are shown in Fig.~\ref{fig:LHCpA}. 
   The ALICE results show $R_{\rm pPb}=1$, constant for $10\leq \pT\leq 50$ GeV/c, while the CMS results agree for $3\leq \pT\leq 20$ GeV/c, with $R_{\rm pPb}=1$, but then show a sharp increase to $R_{\rm pPb}\approx 1.4$ for $40\leq \pT\leq 100$ GeV/c, a jump never before seen in such measurements. For comparison the ATLAS jet measurement at $x_T\geq0.045$ ($\pT\geq113$ GeV/c) (Fig.~\ref{fig:LHCRHICcent}b) is constant at $R_{\rm pPb}\approx 1.2\pm 0.1$. Since there is no $p+p$ comparison measurement for single inclusive particles at \sqs=5.02 TeV, experience suggests that this is the problem, which must be resolved by a high priority \sqs=5.02 TeV $p+p$ comparison run when the LHC starts up again.
   
   There were similar ``exciting results'' at CERN in 1982 which had unexpected consequences. 
\subsection{Experience is the best teacher. Right?}
\begin{figure}[!hbt]
\begin{center}
\raisebox{0.8pc}{\includegraphics[width=0.345\textwidth]{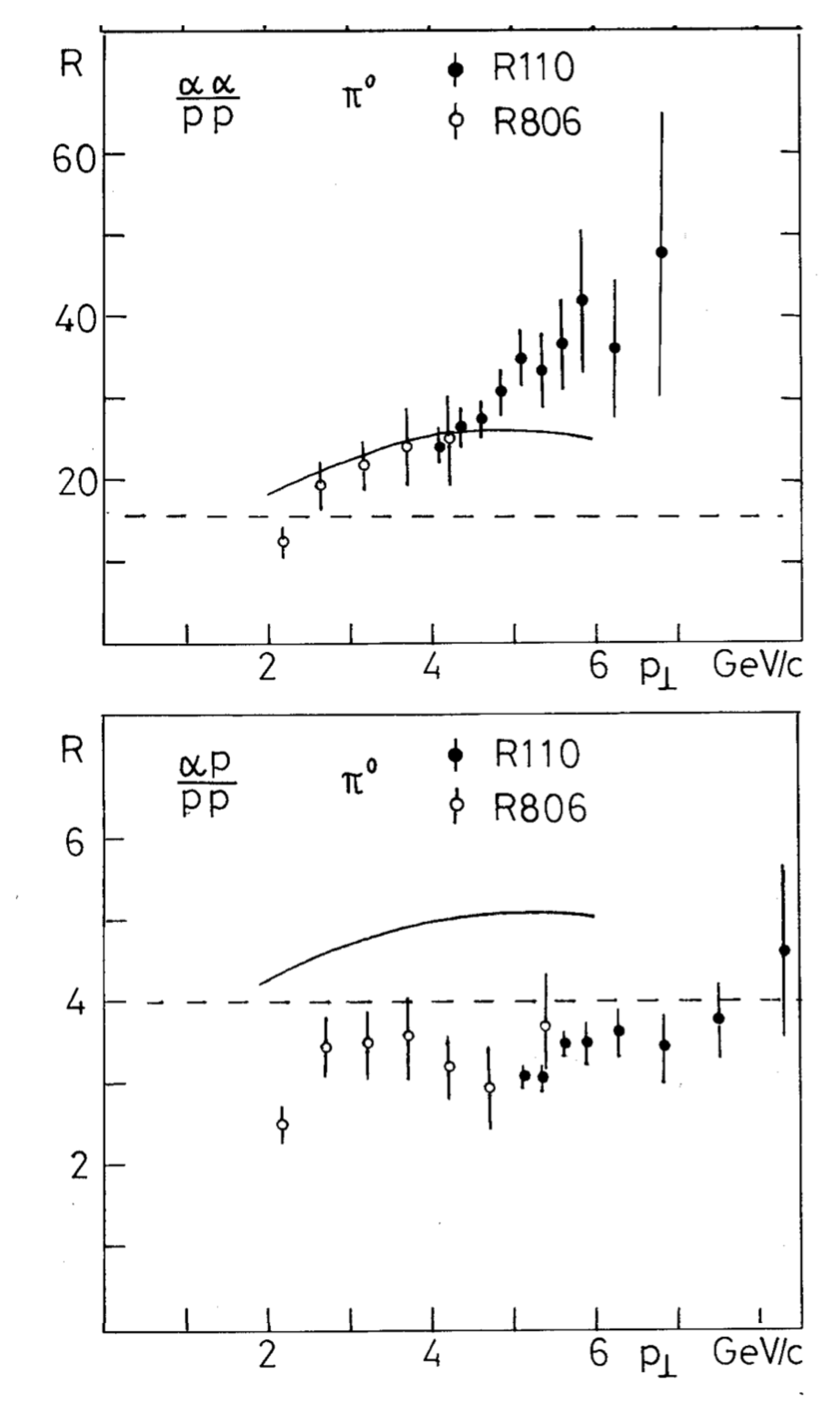}}\hspace*{1.0pc}
\raisebox{0.0pc}{\includegraphics[width=0.45\textwidth]{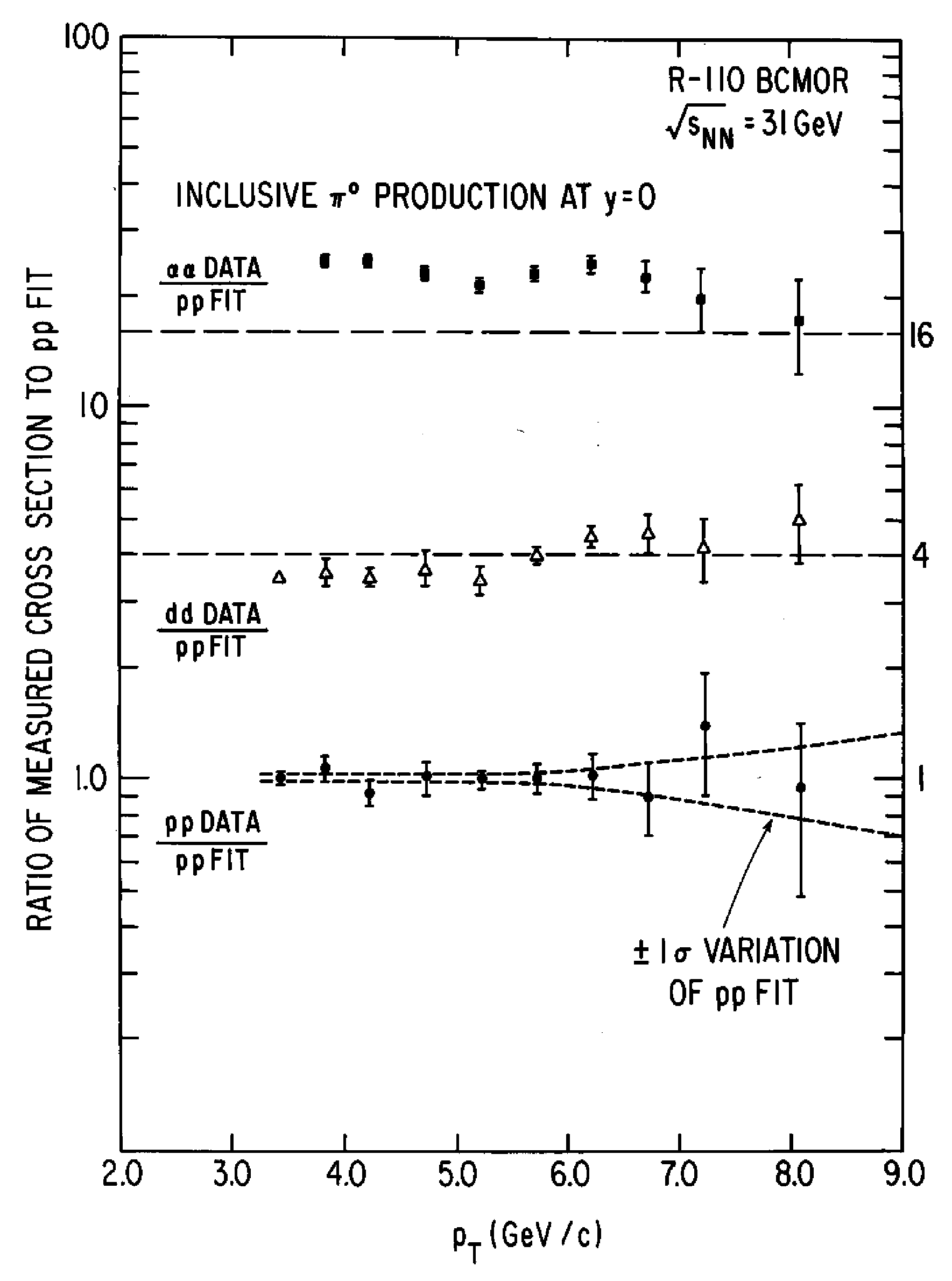}}
\end{center}\vspace*{-1pt}
\caption[]{a) (left) Ratio of cross sections in $\alpha+p$ and $\alpha+\alpha$ interactions to the cross sections in $p+p$  interactions as a function of $p_T$:  $R[(\alpha p\rightarrow \pi^0 +X)/(pp\rightarrow \pi^0 +X)]$ at $\sqrt{s_{NN}}=44$ GeV and $R[(\alpha \alpha\rightarrow \pi^0 +X)/(pp\rightarrow \pi^0 +X)]$ at $\sqrt{s_{NN}}=31$ GeV~\cite{CORPLB116},  compiled by Faessler~\cite{FaesslerPhysinColl}. b) (right) BCMOR measurements~\cite{BCMORPLB185} of the inclusive $\pi^0$ cross sections in $\alpha+\alpha$, $d+d$ and $p+p$ collisions at $\sqrt{s_{NN}}=31$ GeV divided by a fit to the $p+p$ data. }
\label{fig:ExcitingResults}\vspace*{-1pt}
\end{figure}

In 1984, a program of Heavy Ions in the CERN-SPS was approved by the DG, Herwig Schopper, partly due to some ``exciting results'' from $\alpha+\alpha$ collisions in the CERN-ISR (Fig~\ref{fig:ExcitingResults}a)~\cite{FaesslerPhysinColl}.
The large value of the $\alpha\alpha/pp$ cross sections in Fig.~\ref{fig:ExcitingResults}a was WRONG because of an incorrect extrapolation of $p+p$ measurements from \sqs=62.4 to 31 GeV. I complained about this but I was too busy making magnets at ISABELLE at that time---a lucky break in retrospect. Also, because ISABELLE was cancelled in 1983 and the chair of my department, Arthur Schwartzschild, was a nuclear physicist who had heard of this ``exciting result'' by the grapevine and wanted to get collider experience for the RHIC proposal, he offered me a small group of nuclear physicists to participate in the 1983 CERN-ISR $p+p$, $d+d$ and $\alpha+\alpha$ run at\sqsn=31 GeV  (the BCMOR collaboration where B stands for Brookhaven). The correct results are shown in Fig.~\ref{fig:ExcitingResults}b~\cite{BCMORPLB185}. 

This shows that sometimes WRONG RESULTS can have a bigger impact than correct results because they are EXCITING; but this does not excuse making mistakes.
\ack
I would like to acknowledge the fantastic effort by the staff of SUBATECH, the conference organizers, especially Magali Estienne, and the Nantes Police, in finding and returning my wallet totally intact with cash, credit cards and passport, which I had lost leaving the Workshop.
\section*{References}
\bibliography{iopart-num-mjt}

\end{document}